# Faster polynomial multiplication over finite fields

David Harvey

School of Mathematics and Statistics
University of New South Wales
Sydney NSW 2052
Australia

*Email:* d.harvey@unsw.edu.au

Joris van der Hoeven[a], Grégoire Lecerf[b]

Laboratoire d'informatique
UMR 7161 CNRS
École polytechnique
91128 Palaiseau Cedex
France

a. *Email:* vdhoeven@lix.polytechnique.fr
b. *Email:* lecerf@lix.polytechnique.fr

Let $p$ be a prime, and let $\mathsf{M}_p(n)$ denote the bit complexity of multiplying two polynomials in $\mathbb{F}_p[X]$ of degree less than $n$. For $n$ large compared to $p$, we establish the bound $\mathsf{M}_p(n) = O(n \log n\, 8^{\log^* n} \log p)$, where $\log^*$ is the iterated logarithm. This is the first known Fürer-type complexity bound for $\mathbb{F}_p[X]$, and improves on the previously best known bound $\mathsf{M}_p(n) = O(n \log n \log \log n \log p)$.

A.C.M. SUBJECT CLASSIFICATION: G.1.0 Computer-arithmetic, F.2.1 Computations in finite fields

A.M.S. SUBJECT CLASSIFICATION: 68W30, 68Q17, 68W40

## 1. Introduction

Given a ring $R$, a basic problem in complexity theory is to find an upper bound for the cost of multiplying two polynomials in $R[X]$ of degree less than $n$. Several complexity models may be considered. In *algebraic complexity* models, such as the straight-line program model [9, Chapter 4], we count the number of ring operations in $R$, denoted $\mathsf{M}_R(n)$. In this model, the best currently known bound $\mathsf{M}_R(n) = O(n \log n \log \log n)$ was obtained by Cantor and Kaltofen [10]. More precisely, their algorithm performs $O(n \log n \log \log n)$ additions and subtractions and $O(n \log n)$ multiplications in $R$. Their bound may be viewed as an algebraic analogue of the Schönhage–Strassen result for integer multiplication [32] and generalises previous work by Schönhage [31]. These algorithms all rely on suitable incarnations of the fast Fourier transform (FFT) [12]. For details we refer the reader to [17, Chapter 8].

In this paper we are mainly interested in the case that $R$ is the finite field $\mathbb{F}_p = \mathbb{Z}/p\mathbb{Z}$ for some prime $p$. The standard *bit complexity* model based on deterministic multitape Turing machines [26] is more realistic in this setting, as it takes into account the dependence on $p$. We write $\mathsf{M}_p(n)$ for the bit complexity of multiplying two polynomials in $\mathbb{F}_p[X]$ of degree less than $n$. Two basic approaches are known for obtaining good asymptotic bounds for $\mathsf{M}_p(n)$.

If $n$ is not too large compared to $p$, then one may use *Kronecker substitution*, which converts the problem to integer multiplication by packing the coefficients into large integers. Let $\mathsf{I}(n)$ denote the bit complexity of $n$-bit integer multiplication. Throughout the paper, we make the customary assumption that $\mathsf{I}(n)/n$ is increasing, and for convenience we define $\mathsf{I}(x) := \mathsf{I}(\lceil x \rceil)$ for $x \in \mathbb{R}$, $x > 0$. According to our recent sharpening [18] of Fürer's bound [15], we may take

$$\mathsf{I}(n) = O(n \log n\, 8^{\log^* n}), \qquad (1.1)$$

where $\log^* x$ denotes the iterated logarithm for $x \in \mathbb{R}$, i.e.,

$$\begin{aligned} \log^* x &:= \min\{k \in \mathbb{N} : \log^{\circ k} x \leqslant 1\}, \\ \log^{\circ k} &:= \underbrace{\log \circ \cdots \circ \log}_{k \times}. \end{aligned} \qquad (1.2)$$

If $\log n = O(\log p)$, then Kronecker substitution yields $\mathsf{M}_p(n) = O(\mathsf{I}(n \log p))$ (see section 2.6), so we have

$$\mathsf{M}_p(n) = O\big(n \log p \log(n \log p)\, 8^{\log^*(n \log p)}\big).$$

However, Kronecker substitution leads to inferior bounds when $n$ is large compared to $p$, due to coefficient growth in the lifted product in $\mathbb{Z}[X]$. In this situation the best existing method is the algebraic Schönhage–Strassen algorithm (i.e., the Cantor–Kaltofen algorithm). In the Turing model this yields the bound

$$\mathsf{M}_p(n) = O(n \log n \log \log n \log p + n \log n\, \mathsf{I}(\log p)).$$

The first term dominates for large enough $n$, say for $\log \log p\, 8^{\log^* p} = O(\log \log n)$, and we get simply

$$\mathsf{M}_p(n) = O(n \log n \log \log n \log p). \tag{1.3}$$

Since Fürer's breakthrough result, there has been a gap between what is known for integer multiplication and for multiplication in $\mathbb{F}_p[X]$. Namely, it has remained an open question whether the factor $\log \log n$ appearing in (1.3) can be improved to $K^{\log^* n}$ for some $K > 1$. It seems that Fürer's construction does not work in $\mathbb{F}_p[X]$.

## 1.1. Our contributions

The main result of this paper is a new algorithm that completely closes the gap mentioned above.

THEOREM 1.1. *The bound*

$$\mathsf{M}_p(n) = O\big(n \log p \log(n \log p)\, 8^{\log^*(n \log p)}\big)$$

*holds uniformly for all $n \geqslant 2$ and all primes $p$.*

The bound follows immediately via Kronecker substitution if $\log n = O(\log p)$. The bulk of the paper concerns the reverse case where $n$ is large compared to $p$. For example, if $\log \log p = O(\log n)$, the bound becomes simply

$$\mathsf{M}_p(n) = O(n \log n\, 8^{\log^* n} \log p).$$

As promised, this replaces the factor $\log \log n$ in (1.3) by $8^{\log^* n}$.

The basic idea of the new algorithm is as follows. We first construct a special extension $\mathbb{F}_{p^\kappa}$ of $\mathbb{F}_p$, where $\kappa$ is exponentially smaller than $n$, and for which $p^\kappa - 1$ has many small divisors. These divisors are themselves exponentially smaller than $n$, yet they are so numerous that their product is comparable to $n$. We now convert the given multiplication problem in $\mathbb{F}_p[X]$ to multiplication in $\mathbb{F}_{p^\kappa}[Y]$, by cutting up the polynomials into small pieces, and we then multiply in $\mathbb{F}_{p^\kappa}[Y]$ by using FFTs over $\mathbb{F}_{p^\kappa}$. Applying the Cooley–Tukey method to the small divisors of $p^\kappa - 1$, we decompose the FFTs into transforms of exponentially shorter lengths (not necessarily powers of two). We use Bluestein's method to convert each short transform to a convolution over $\mathbb{F}_{p^\kappa}$, and then use Kronecker substitution (from bivariate to univariate polynomials) to convert this to multiplication in $\mathbb{F}_p[X]$. These latter multiplications are then handled recursively. We continue the recursion until $n$ is comparable to $p$, at which point we switch to using Kronecker substitution (from polynomials to integers).

In many respects this approach is very similar to the algorithm for integer multiplication introduced in [18], but there are various additional technical issues to address. We recommend [18] as a gentle introduction to the main ideas.

We also prove the following conditional result.

THEOREM 1.2. *Assume Conjectures 8.1, 8.3 and 8.5. Then the bound*

$$\mathsf{M}_p(n) = O\big(n \log p \log(n \log p)\, 4^{\log^*(n \log p)}\big)$$

*holds uniformly for all $n \geqslant 2$ and all primes $p$.*

Of these conjectures, Conjecture 8.3 seems to be in reach for specialists in analytic number theory; Conjecture 8.5 is at least as hard as Artin's conjecture on primitive roots, but may be tractable under the generalised Riemann hypothesis (GRH); and Conjecture 8.1 is almost as strong as the Lenstra–Pomerance–Wagstaff conjecture on the distribution of Mersenne primes, and seems to be inaccessible at present.

## 1.2. Outline of the paper

The paper is structured as follows. In section 2, we start with a survey of relevant basic techniques: discrete Fourier transforms (DFTs) and FFTs, Bluestein's chirp transform, and Kronecker substitution. Most of this material is repeated from [18, Section 2] for the convenience of the reader; however, sections 2.5 and 2.6 differ substantially.

In section 3, we recall basic complexity results for arithmetic in finite fields. In particular, we consider the construction of irreducible polynomials in $\mathbb{F}_p[X]$, algorithms for finding roots of unity in $\mathbb{F}_{p^\kappa}$, and the cost of arithmetic in $\mathbb{F}_{p^\kappa}$ and $\mathbb{F}_{p^\kappa}[Y]$.

In section 4, we show how to construct special extensions of $\mathbb{F}_p$ whose multiplicative group has a large subgroup of highly smooth order, i.e., is divisible by many small integers. The main tool is [2, Theorem 3].

Section 5 gives complexity bounds for functions that satisfy recurrence inequalities involving postcompositions with "logarithmically slow" functions. The prototype of such an inequality is $T(n) \leqslant K T(\log n) + L$, where $K$ and $L$ are constants. The definitions and theorems are duplicated from [18, Section 5]; for the proofs, see [18].

To minimise the constant $K$ in the bound $\mathsf{M}_p(n) = O(n \log n\, K^{\log^* n} \log p)$ (for $n$ large relative to $p$), we need one more tool: in section 6, we present a polynomial analogue of the Crandall–Fagin convolution algorithm [13]. This allows us to convert a cyclic convolution over $\mathbb{F}_p$ of length $n$, where $n$ is arbitrary, into a cyclic convolution over $\mathbb{F}_{p^\kappa}$ of somewhat smaller length $N$, where $N$ is prescribed and where $\kappa \approx 2 n / N$. (We can still obtain $K = 16$ without using this Crandall–Fagin analogue, but we do not know how to reach $K = 8$ without it.)

Section 7 is devoted to the proof of Theorem 1.1. Section 8 gives the proof of Theorem 1.2, and discusses the three number-theoretic hypotheses of that theorem.

The last section offers some final notes and suggested directions for generalisation. We first quickly dispense with the bit complexity of multiplication in $\mathbb{F}_q[X]$ for prime powers $q$, and in $(\mathbb{Z}/m\mathbb{Z})[X]$ for arbitrary integers $m \geqslant 1$. We then sketch some algebraic complexity bounds for polynomial multiplication over $\mathbb{F}_p$-algebras and $(\mathbb{Z}/m\mathbb{Z})$-algebras, especially in the straight-line program model. Our techniques also give rise to new strategies for polynomial evaluation-interpolation over $\mathbb{F}_q$. This may for instance be applied to the efficient multiplication of polynomial matrices over $\mathbb{F}_q$. Although we have not implemented any of our algorithms yet, we conclude by a few remarks on their expected efficiency in practice.

**Notations.** We use Hardy's notations $f \prec g$ for $f = o(g)$, and $f \asymp g$ for $f = O(g)$ and $g = O(f)$. The "soft-Oh" notation $f(n) = \tilde{O}(g(n))$ means that $f(n) = g(n)\, (\log(3 + g(n)))^{O(1)}$ (see [17, Chapter 25, Section 7] for details). The symbol $\mathbb{R}^{\geqslant}$ denotes the set of non-negative real numbers, and $\mathbb{N}$ denotes $\{0, 1, 2, \ldots\}$. For a ring $R$ and $n \in \mathbb{N}$, we write $R[X]_n := \{P \in R[X] : \deg P < n\}$.

We will write $\lg n := \lceil \log n / \log 2 \rceil$. In expressions like $\log \log p$ or $\lg \lg p$, we tacitly assume that the function is adjusted so as to take positive values for small primes such as $p = 2$.

**Acknowledgments.** We would like to thank Karim Belabas for directing us to the paper [2], and Igor Shparlinski for answering some analytic number theory questions.

## 2. SURVEY OF CLASSICAL TOOLS

This section recalls basic facts on Fourier transforms and related techniques used in subsequent sections. For more details and historical references we refer the reader to standard books on the subject such as [3, 9, 17, 30].

### 2.1. Arrays and sorting

In the Turing model, we have available a fixed number of linear tapes. An $n_1 \times \cdots \times n_d$ array $M_{i_1,\ldots,i_d}$ of $b$-bit elements is stored as a linear array of $n_1 \cdots n_d b$ bits. We generally assume that the elements are ordered lexicographically by $(i_1, \ldots, i_d)$, though this is just an implementation detail.

What is significant from a complexity point of view is that occasionally we must switch representations, to access an array (say 2-dimensional) by "rows" or by "columns". In the Turing model, we may transpose an $n_1 \times n_2$ matrix of $b$-bit elements in time $O(b\, n_1 n_2 \lg \min(n_1, n_2))$, using the algorithm of [7, Appendix]. Briefly, the idea is to split the matrix into two halves along the "short" dimension, and transpose each half recursively.

We will also require more complex rearrangements of data, for which we resort to sorting. Suppose that $X$ is a totally ordered set, whose elements are represented by bit strings of length $b$, and suppose that we can compare elements of $X$ in time $O(b)$. Then an array of $n$ elements of $X$ may be sorted in time $O(b\, n \lg n)$ using merge sort [22], which can be implemented efficiently on a Turing machine.

## 2.2. Discrete Fourier transforms

Let $R$ be a commutative ring with identity and let $n \geqslant 1$. An element $\omega \in R$ is said to be a *principal $n$-th root of unity* if $\omega^n = 1$ and

$$\sum_{k=0}^{n-1} (\omega^i)^k = 0 \tag{2.1}$$

for all $i \in \{1, ..., n-1\}$. In this case, we define the *discrete Fourier transform* (or DFT) of an $n$-tuple $a = (a_0, ..., a_{n-1}) \in R^n$ with respect to $\omega$ to be $\mathrm{DFT}_\omega(a) = \hat{a} = (\hat{a}_0, ..., \hat{a}_{n-1}) \in R^n$ where

$$\hat{a}_i := a_0 + a_1 \omega^i + \cdots + a_{n-1} \omega^{(n-1)i}.$$

That is, $\hat{a}_i$ is the evaluation of the polynomial $A(X) := a_0 + a_1 X + \cdots + a_{n-1} X^{n-1}$ at $\omega^i$.

If $\omega$ is a principal $n$-th root of unity, then so is its inverse $\omega^{-1} = \omega^{n-1}$, and we have

$$\mathrm{DFT}_{\omega^{-1}}(\mathrm{DFT}_\omega(a)) = n\, a.$$

Indeed, writing $b := \mathrm{DFT}_{\omega^{-1}}(\mathrm{DFT}_\omega(a))$, the relation (2.1) implies that

$$b_i = \sum_{j=0}^{n-1} \hat{a}_j \omega^{-ji} = \sum_{j=0}^{n-1} \sum_{k=0}^{n-1} a_k \omega^{j(k-i)} = \sum_{k=0}^{n-1} a_k \sum_{j=0}^{n-1} \omega^{j(k-i)} = \sum_{k=0}^{n-1} a_k (n\, \delta_{i,k}) = n\, a_i,$$

where $\delta_{i,k} = 1$ if $i = k$ and $\delta_{i,k} = 0$ otherwise.

**Remark 2.1.** In all of the new algorithms introduced in this paper, we actually work over a field, whose characteristic does not divide $n$. In this setting, the concept of principal root of unity coincides with the more familiar *primitive root of unity*.

## 2.3. The Cooley–Tukey FFT

Let $\omega$ be a principal $n$-th root of unity and let $n = n_1 n_2$ where $1 < n_1 < n$. Then $\omega^{n_1}$ is a principal $n_2$-th root of unity and $\omega^{n_2}$ is a principal $n_1$-th root of unity. Moreover, for any $i_1 \in \{0, ..., n_1 - 1\}$ and $i_2 \in \{0, ..., n_2 - 1\}$, we have

$$\begin{aligned}\hat{a}_{i_1 n_2 + i_2} &= \sum_{k_1=0}^{n_1-1} \sum_{k_2=0}^{n_2-1} a_{k_2 n_1 + k_1} \omega^{(k_2 n_1 + k_1)(i_1 n_2 + i_2)} \\ &= \sum_{k_1=0}^{n_1-1} \omega^{k_1 i_2} \left( \sum_{k_2=0}^{n_2-1} a_{k_2 n_1 + k_1} (\omega^{n_1})^{k_2 i_2} \right) (\omega^{n_2})^{k_1 i_1}. \end{aligned} \tag{2.2}$$

If $\mathcal{A}_1$ and $\mathcal{A}_2$ are algorithms for computing DFTs of length $n_1$ and $n_2$, we may use (2.2) to construct an algorithm $\mathcal{A}_1 \odot \mathcal{A}_2$ for computing DFTs of length $n$ as follows.

For each $k_1 \in \{0, ..., n_1 - 1\}$, the sum inside the brackets corresponds to the $i_2$-th coefficient of a DFT of the $n_2$-tuple $(a_{0 n_1 + k_1}, ..., a_{(n_2-1)n_1 + k_1}) \in R^{n_2}$ with respect to $\omega^{n_1}$. Evaluating these *inner DFTs* requires $n_1$ calls to $\mathcal{A}_2$. Next, we multiply by the *twiddle factors* $\omega^{k_1 i_2}$, at a cost of $n$ operations in $R$. (Actually, fewer than $n$ multiplications are required, as some of the twiddle factors are equal to 1. This optimisation, while important in practice, has no asymptotic effect on the algorithms discussed in this paper.) Finally, for each $i_2 \in \{0, ..., n_2 - 1\}$, the outer sum corresponds to the $i_1$-th coefficient of a DFT of an $n_1$-tuple in $R^{n_1}$ with respect to $\omega^{n_2}$. These *outer DFTs* require $n_2$ calls to $\mathcal{A}_1$.

Denoting by $\mathsf{F}_R(n)$ the number of ring operations needed to compute a DFT of length $n$, and assuming that we have available a precomputed table of twiddle factors, we obtain

$$\mathsf{F}_R(n_1 n_2) \leqslant n_1 \mathsf{F}_R(n_2) + n_2 \mathsf{F}_R(n_1) + n.$$

For a factorisation $n = n_1 \cdots n_d$, this yields recursively

$$\mathsf{F}_R(n) \leqslant \sum_{i=1}^{d} \frac{n}{n_i} \mathsf{F}_R(n_i) + (d-1) n. \tag{2.3}$$

The corresponding algorithm is denoted $\mathcal{A}_1 \odot \cdots \odot \mathcal{A}_d$. The $\odot$ operation is neither commutative nor associative; the above expression will always be taken to mean $(\cdots((\mathcal{A}_1 \odot \mathcal{A}_2) \odot \mathcal{A}_3) \odot \cdots) \odot \mathcal{A}_d$.

The above discussion requires several modifications in the Turing model. Assume that elements of $R$ are represented by $b$ bits.

First, for $\mathcal{A}_1 \odot \mathcal{A}_2$, we must add a rearrangement cost of $O(b n \lg \min(n_1, n_2))$ to efficiently access the rows and columns for the recursive subtransforms (see section 2.1). For the general case $\mathcal{A}_1 \odot \cdots \odot \mathcal{A}_d$, the total rearrangement cost is bounded by $O(\sum_i b n \lg n_i) = O(b n \lg n)$.

Second, we will sometimes use *non-algebraic* algorithms to compute the subtransforms, so it may not make sense to express their cost in terms of $\mathsf{F}_R$. The relation (2.3) therefore becomes

$$\mathsf{F}(n) \leqslant \sum_{i=1}^{d} \frac{n}{n_i} \mathsf{F}(n_i) + (d-1) n \, \mathsf{m}_R + O(b n \lg n), \tag{2.4}$$

where $\mathsf{F}(n)$ is the (Turing) cost of a transform of length $n$ over $R$, and where $\mathsf{m}_R$ is the cost of a single multiplication in $R$.

Finally, we point out that $\mathcal{A}_1 \odot \mathcal{A}_2$ requires access to a table of twiddle factors $\omega^{i_1 i_2}$, ordered lexicographically by $(i_1, i_2)$, for $0 \leqslant i_1 < n_1$, $0 \leqslant i_2 < n_2$. Assuming that we are given as input a precomputed table of the form $1, \omega, \ldots, \omega^{n-1}$, we must show how to extract the required twiddle factor table in the correct order. We first construct a list of triples $(i_1, i_2, i_1 i_2)$, ordered by $(i_1, i_2)$, in time $O(n \lg n)$; then sort by $i_1 i_2$ in time $O(n \lg^2 n)$ (see section 2.1); then merge with the given root table to obtain a table $(i_1, i_2, \omega^{i_1 i_2})$, ordered by $i_1 i_2$, in time $O(n(b + \lg n))$; and finally sort by $(i_1, i_2)$ in time $O(n \lg n (b + \lg n))$. The total cost of the extraction is thus $O(n \lg n (b + \lg n))$.

The corresponding cost for $\mathcal{A}_1 \odot \cdots \odot \mathcal{A}_d$ is determined as follows. Assuming that the table $1, \omega, \ldots, \omega^{n-1}$ is given as input, we first extract the subtables of $(n_1 \cdots n_i)$-th roots of unity for $i = d-1, \ldots, 2$ in time $O((n_1 \cdots n_d + \cdots + n_1 n_2)(b + \lg n)) = O(n(b + \lg n))$. Extracting the twiddle factor table for the decomposition $(n_1 \cdots n_{i-1}) \times n_i$ then costs $O(n_1 \cdots n_i \lg n (b + \lg n))$; the total over all $i$ is again $O(n \lg n (b + \lg n))$.

**Remark 2.2.** An alternative approach is to compute the twiddle factors directly in the correct order. When working over $\mathbb{C}$, as in [18, Section 3], this requires a slight increase in the working precision. Similar comments apply to the root tables used in Bluestein's algorithm in section 2.5.

## 2.4. Fast Fourier multiplication

Let $\omega$ be a principal $n$-th root of unity in $R$ and assume that $n$ is invertible in $R$. Consider two polynomials $A = a_0 + \cdots + a_{n-1} X^{n-1}$ and $B = b_0 + \cdots + b_{n-1} X^{n-1}$ in $R[X]$. Let $C = c_0 + \cdots + c_{n-1} X^{n-1}$ be the polynomial defined by

$$c := \tfrac{1}{n} \mathrm{DFT}_{\omega^{-1}}(\mathrm{DFT}_\omega(a) \, \mathrm{DFT}_\omega(b)),$$

where the product of the DFTs is taken pointwise. By construction, we have $\hat{c} = \hat{a} \, \hat{b}$, which means that $C(\omega^i) = A(\omega^i) B(\omega^i)$ for all $i \in \{0, \ldots, n-1\}$. The product $S = s_0 + \cdots + s_{n-1} X^{n-1}$ of $A$ and $B$ modulo $X^n - 1$ also satisfies $S(\omega^i) = A(\omega^i) B(\omega^i)$ for all $i$. Consequently, $\hat{s} = \hat{a} \, \hat{b}$, $s = \mathrm{DFT}_{\omega^{-1}}(\hat{s})/n = c$, whence $C = S$.

For polynomials $A, B \in R[X]$ with $\deg A < n$ and $\deg B < n$, we thus obtain an algorithm for the computation of $A B$ modulo $X^n - 1$ using at most $3 \mathsf{F}_R(n) + O(n)$ operations in $R$. Modular products of this type are also called *cyclic convolutions*. If $\deg (A B) < n$, then we may recover the product $A B$ from its reduction modulo $X^n - 1$. This multiplication method is called *FFT multiplication*.

If one of the arguments (say $B$) is fixed and we want to compute many products $AB$ (or cyclic convolutions) for different $A$, then we may precompute $\text{DFT}_\omega(b)$, after which each new product $AB$ can be computed using only $2\,\mathsf{F}_R(n)+O(n)$ operations in $R$.

## 2.5. Bluestein's chirp transform

We have shown above how to multiply polynomials using DFTs. Inversely, it is possible to reduce the computation of DFTs — of arbitrary length, not necessarily a power of two — to polynomial multiplication [5], as follows.

Let $\omega$ be a principal $n$-th root of unity and consider an $n$-tuple $a \in R^n$. First consider the case that $n$ is odd. Let
$$f_i := \omega^{(i^2-i)/2}, \quad f'_i := \omega^{(i^2+i)/2}, \quad g_i := \omega^{(-i^2-i)/2}.$$

Note that $(i^2-i)/2$ and $(i^2+i)/2$ are integers, and that
$$g_{i+n} = \omega^{(-(i+n)^2-(i+n))/2} = \omega^{(-i^2-i)/2}\,\omega^{n(-i-(n+1)/2)} = g_i.$$

Let $F := f_0\,a_0 + \cdots + f_{n-1}\,a_{n-1}\,X^{n-1}$, $G := g_0 + \cdots + g_{n-1}\,X^{n-1}$ and $C := c_0 + \cdots + c_{n-1}\,X^{n-1} \equiv FG$ modulo $X^n - 1$. Then
$$f'_i\,c_i = \sum_{j=0}^{n-1} f'_i\,(f_j\,a_j)\,g_{i-j} = \sum_{j=0}^{n-1} \omega^{(i^2+i)/2}\,\omega^{(j^2-j)/2}\,\omega^{(-(i-j)^2-(i-j))/2}\,a_j = \sum_{j=0}^{n-1} \omega^{ij}\,a_j = \hat{a}_i.$$

In other words, the computation of a DFT of odd length $n$ reduces to a cyclic convolution product of the same length $n$, together with $O(n)$ additional operations in $R$. Notice that the polynomial $G$ is fixed and independent of $a$ in this product.

Now suppose that $n$ is even. In this case we require that 2 be invertible in $R$, and that $\omega^{n/2} = -1$. Let $\sigma := (-1)^{n/2}$, and put
$$f_i := \omega^{i^2}, \quad f'_i := \omega^{i^2+i}, \quad g_i := \omega^{-i^2} + \omega^{-i^2-i}.$$

Then
$$g_{i+\frac{n}{2}} = \sigma\left(\omega^{-i^2} - \omega^{-i^2-i}\right), \quad g_{i+n} = g_i,$$
and
$$\begin{aligned}\tfrac{1}{2}\left(g_i + \sigma\,g_{i+\frac{n}{2}}\right) &= \omega^{-i^2}, \\ \tfrac{1}{2}\left(g_i - \sigma\,g_{i+\frac{n}{2}}\right) &= \omega^{-i^2-i}.\end{aligned}$$

As above, let $F := f_0\,a_0 + \cdots + f_{n-1}\,a_{n-1}\,X^{n-1}$, $G := g_0 + \cdots + g_{n-1}\,X^{n-1}$ and $C := c_0 + \cdots + c_{n-1}\,X^{n-1} \equiv FG$ modulo $X^n - 1$. Then for $i \in \{0,\ldots,\tfrac{n}{2}-1\}$ we have
$$\begin{aligned}\tfrac{1}{2} f_i\left(c_i + \sigma\,c_{i+\frac{n}{2}}\right) &= \tfrac{1}{2} f_i \sum_{j=0}^{n-1} f_j\,a_j\left(g_{i-j} + \sigma\,g_{i-j+\frac{n}{2}}\right) \\ &= \sum_{j=0}^{n-1} \omega^{i^2}\,\omega^{j^2}\,\omega^{-(i-j)^2}\,a_j = \sum_{j=0}^{n-1} \omega^{2ij}\,a_j = \hat{a}_{2i},\end{aligned}$$
and similarly
$$\tfrac{1}{2} f'_i\left(c_i - \sigma\,c_{i+\frac{n}{2}}\right) = \sum_{j=0}^{n-1} \omega^{i^2+i}\,\omega^{j^2}\,\omega^{-(i-j)^2-(i-j)}\,a_j = \sum_{j=0}^{n-1} \omega^{(2i+1)j}\,a_j = \hat{a}_{2i+1}.$$

Again, the DFT reduces to a convolution of length $n$, together with $O(n)$ additional operations in $R$.

The only complication in the Turing model is the cost of extracting the $f_i$, $f'_i$ and $g_i$ in the correct order. For example, consider the $f_i$ in the case that $n$ is odd. Given as input a precomputed table $1, \omega, \omega^2, \ldots, \omega^{n-1}$, we may extract the $f_i$ in time $O(n \lg n\,(b + \lg n))$ by applying the strategy from section 2.3 to the pairs $(i, \tfrac{1}{2}(i^2-i) \bmod n)$ for $0 \leqslant i < n$. The other sequences are handled similarly. For the $g_i$ in the case that $n$ is even, we also need to perform $O(n)$ additions in $R$; assuming that additions in $R$ have cost $O(b)$, this is already covered by the above bound.

**Remark 2.3.** Bluestein's original formulation used the weights $f_i := \omega^{i^2/2}$, $g_i := \omega^{-i^2/2}$. This has two drawbacks in our setting. First, it requires the presence of $2n$-th roots of unity, which may not exist in $R$. Second, if $n$ is odd, it leads to negacyclic convolutions, rather than cyclic convolutions. The variants used here avoid both of these problems.

**Remark 2.4.** It is also possible to give variants of the new multiplication algorithms in which Bluestein's transform is replaced by a different method for converting DFTs to convolutions, such as Rader's algorithm [29].

## 2.6. Kronecker substitution

Multiplication in $\mathbb{F}_p[X]$ may be reduced to multiplication in $\mathbb{Z}$ using the classical technique of *Kronecker substitution* [17, Corollary 8.27]. More precisely, let $n > 0$ and suppose that we are given two polynomials $A, B \in \mathbb{F}_p[X]$ of degree less than $n$. Let $\tilde{A}, \tilde{B} \in \mathbb{Z}[X]$ be lifts of these polynomials, with coefficients $A_i$ and $B_i$ satisfying $0 \leqslant A_i, B_i < p$. Then for the product $\tilde{C} = \tilde{A}\tilde{B}$ we have $0 \leqslant C_i < np^2$, and the coefficients of $\tilde{C}$ may be read off the integer product $\tilde{C}(2^N) = \tilde{A}(2^N)\tilde{B}(2^N)$ where $N := 2\lg p + \lg n$. We deduce the coefficients of $C := AB$ by dividing each $C_i$ by $p$. This shows that $\mathsf{M}_p(n) = O(\mathsf{I}(n(\lg p + \lg n)) + n\,\mathsf{I}(\lg p))$. Using the assumption that $\mathsf{I}(n)/n$ is increasing, we obtain

$$\mathsf{M}_p(n) = O(\mathsf{I}(n(\lg p + \lg n))).$$

If we also assume that $\lg n = O(\lg p)$, i.e., that the degree is not too large, then this becomes simply

$$\mathsf{M}_p(n) = O(\mathsf{I}(n \lg p)). \tag{2.5}$$

A second type of Kronecker substitution reduces bivariate polynomial multiplication to the univariate case. Indeed, let $n \geqslant 1$ and $\kappa \geqslant 1$, and suppose that $A, B \in \mathbb{F}_p[X, Z]$, where $\deg_X A$, $\deg_X B < n$ and $\deg_Z A, \deg_Z B < \kappa$. We may then recover $C := AB$ from the univariate product $C(Y^{2\kappa}, Y) = A(Y^{2\kappa}, Y) B(Y^{2\kappa}, Y)$ in $\mathbb{F}_p[Y]$. Note that $A(Y^{2\kappa}, Y)$ and $B(Y^{2\kappa}, Y)$ have degree less than $2n\kappa$, so the cost of the bivariate product is $\mathsf{M}_p(2n\kappa) + O(n\kappa \lg p)$.

The same method works for computing cyclic convolutions: to multiply $A, B \in \mathbb{F}_p[X, Z] / (X^n - 1)$, the same substitution leads to a product in $\mathbb{F}_p[Y] / (Y^{2n\kappa} - 1)$. The cost is thus $\mathsf{M}'_p(2n\kappa) + O(n\kappa \lg p)$, where $\mathsf{M}'_p(d)$ denotes the cost of a multiplication in $\mathbb{F}_p[X]/(X^d - 1)$.

## 3. Arithmetic in finite fields

Let $p$ be a prime and let $\kappa \geqslant 1$. In this section we review basic results concerning arithmetic in $\mathbb{F}_{p^\kappa}$ and $\mathbb{F}_{p^\kappa}[Y]$.

We assume throughout that $\mathbb{F}_{p^\kappa}$ is represented as $\mathbb{F}_p[Z]/P$ for some irreducible monic polynomial $P \in \mathbb{F}_p[Z]$ of degree $\kappa$. Thus an element of $\mathbb{F}_{p^\kappa}$ is represented uniquely by its residue modulo $P$, i.e., by a polynomial $F \in \mathbb{F}_p[Z]$ of degree less than $\kappa$.

LEMMA 3.1. *Let $p$ be a prime and let $\kappa \geqslant 1$. We may compute a monic irreducible polynomial $P \in \mathbb{F}_p[Z]$ of degree $\kappa$ in time $\tilde{O}(\kappa^4 p^{1/2})$.*

**Proof.** See [33, Theorem 3.2]. □

The above complexity bound is very pessimistic in practice. Better complexity bounds are known if we allow randomised algorithms, or unproved hypotheses. For instance, assuming GRH, the bound reduces to $(\kappa \log p)^{O(1)}$ [1].

We now consider the cost of arithmetic in $\mathbb{F}_{p^\kappa}$, assuming that $P$ is given. Addition and subtraction in $\mathbb{F}_{p^\kappa}$ have cost $O(\kappa \lg p)$. For multiplication we will always use the Schönhage–Strassen algorithm. Denote by $\mathsf{S}_p(\kappa)$ the cost of multiplying polynomials in $\mathbb{F}_p[Z]_\kappa$ by this method, i.e., using the polynomial variant [31] for the polynomials themselves, followed by the integer version [32] to handle the coefficient multiplications. Then we have

$$\mathsf{S}_p(\kappa) = O(\kappa \lg \kappa \lg \lg \kappa \lg p \lg \lg p \lg \lg \lg p).$$

Of course, we could use the new multiplication algorithm recursively for these products, but it turns out that Schönhage–Strassen is fast enough, and leads to a simpler complexity analysis in section 7. Let $\mathsf{D}_p(\kappa)$ denote the cost of dividing a polynomial in $\mathbb{F}_p[Z]_{2\kappa}$ by $P$, returning the quotient and remainder. Using Newton's method [17, Chapter 9] we have $\mathsf{D}_p(\kappa) = O(\mathsf{S}_p(\kappa))$. Thus elements of $\mathbb{F}_{p^\kappa}$ may be multiplied in time $O(\mathsf{S}_p(\kappa))$.

Let $N \geqslant 1$ be a divisor of $p^\kappa - 1$. To compute a DFT over $\mathbb{F}_{p^\kappa}$ of length $N$, we must first have access to a primitive $N$-th root of unity in $\mathbb{F}_{p^\kappa}$. In general it is very difficult to find a primitive root of $\mathbb{F}_{p^\kappa}$, as it requires knowledge of the factorisation of $p^\kappa - 1$. However, to find a primitive $N$-th root, Lemma 3.3 below shows that it is enough to know the factorisation of $N$. The construction relies on the following existence result.

LEMMA 3.2. *Let $\ell$ be a positive integer such that $p^\ell > c_1 r^4 (\log r + 1)^4 \kappa^2$, where $r$ is the number of distinct prime divisors of $p^\kappa - 1$, and where $c_1 > 0$ is a certain absolute constant. Then there exists a monic irreducible polynomial $\Theta \in \mathbb{F}_p[Z]$ of degree $\ell$ such that $\Theta$ modulo $P$ is a primitive root of unity for $\mathbb{F}_{p^\kappa} = \mathbb{F}_p[Z]/P$.*

**Proof.** This is [34, Theorem 1.1]. □

LEMMA 3.3. *Assume that $N$ divides $p^\kappa - 1$ and that the factorisation of $N$ is given. Then we may compute a primitive $N$-th root of unity in $\mathbb{F}_p[Z]/P$ in time $\tilde{O}(\kappa^9 p)$.*

**Proof.** Testing whether a given $\alpha \in \mathbb{F}_{p^\kappa}$ is a primitive $N$-th root of unity reduces to checking that $\alpha^N = 1$ and $\alpha^{N/s} \neq 1$ for every prime divisor $s$ of $N$. According to Lemma 3.2, it suffices to apply this test to $\alpha := (\Theta \bmod P)^{(p^\kappa - 1)/N}$ for $\Theta \in \mathbb{F}_p[Z]$ running over all monic polynomials of degree $\ell := \lceil \log(c_1 r^4 (\log r + 1)^4 \kappa^2)/\log p \rceil$, where $r = O(\kappa \lg p)$ is as in the lemma. The number of candidates is at most $p^\ell \leqslant c r^4 (\log r + 1)^4 \kappa^2 p = \tilde{O}(\kappa^6 p)$, and each candidate can be tested in time $\tilde{O}(r \lg N (\kappa \lg p)) = \tilde{O}(\kappa^3 \lg^3 p)$ using binary exponentiation. □

Of course, we can do much better if randomised algorithms are allowed, since $\zeta^{(p^\kappa - 1)/N}$ is reasonably likely to be a primitive $N$-th root for randomly selected $\zeta$.

Finally, we consider polynomial multiplication over $\mathbb{F}_{p^\kappa}$. Let $n \geqslant 1$, and let $A, B \in \mathbb{F}_{p^\kappa}[X]_n$. Let $\tilde{A}, \tilde{B} \in \mathbb{F}_p[X, Z]$ be their natural lifts, i.e., of degree less than $\kappa$ with respect to $Z$. The bivariate product $\tilde{C} := \tilde{A}\tilde{B}$ may be computed using Kronecker substitution (section 2.6) in time $\mathsf{M}_p(2n\kappa) + O(n\kappa \lg p)$. Writing $\tilde{C} = \sum_{i=0}^{2n-2} \tilde{C}_i(Z) X^i$, to recover $AB$ we must divide each $\tilde{C}_i$ by $P$. Denoting by $\mathsf{M}_{p^\kappa}(n)$ the complexity of the original multiplication problem, we obtain

$$\begin{aligned} \mathsf{M}_{p^\kappa}(n) &= \mathsf{M}_p(2n\kappa) + O(n\mathsf{D}_p(\kappa)) \\ &= \mathsf{M}_p(2n\kappa) + O(n\mathsf{S}_p(\kappa)). \end{aligned}$$

As in section 2.6, the same method may be used for cyclic convolutions. If $\mathsf{M}'_{p^\kappa}(n)$ denotes the cost of multiplication in $\mathbb{F}_{p^\kappa}[X]/(X^n - 1)$, then we get

$$\mathsf{M}'_{p^\kappa}(n) = \mathsf{M}'_p(2n\kappa) + O(n\mathsf{S}_p(\kappa)).$$

**Remark 3.4.** In the other direction, we can also reduce multiplication in $\mathbb{F}_p[X]$ to multiplication in $\mathbb{F}_{p^\kappa}[Y]$ by splitting the inputs into chunks of size $\lfloor \kappa/2 \rfloor$. This leads to the bound $\mathsf{M}_p(n) \leqslant \mathsf{M}_{p^\kappa}(\lceil n/\lfloor \kappa/2 \rfloor \rceil) + O(n \lg p)$ for $n \geqslant \kappa$. A variant of this procedure is developed in detail in section 6.

## 4. Preparing for DFTs of large smooth orders

The aim of this section is to prove the following theorem, which allows us to construct "small" extensions $\mathbb{F}_{p^\lambda}$ of $\mathbb{F}_p$ containing "many" roots of unity of "low" order. Recall that a positive integer is said to be *y-smooth* if all of its prime divisors are less than or equal to $y$.

THEOREM 4.1. *There exist computable absolute constants $c_3 > c_2 > 0$ and $n_0 \in \mathbb{N}$ with the following properties. Let $p$ be a prime and let $n \geqslant n_0$. Then there exists an integer $\lambda$ in the interval*

$$(\lg n)^{c_2 \lg \lg \lg n} < \lambda < (\lg n)^{c_3 \lg \lg \lg n},$$

and a $(\lambda+1)$-smooth integer $M \geqslant n$, such that $M \mid p^\lambda - 1$. Moreover, given $p$ and $n$, we may compute $\lambda$ and the prime factorisation of $M$ in time $O((\lg n)^{\lg \lg n})$.

The proof depends on a series of preparatory lemmas. For $\lambda \geqslant 2$, define

$$H_\lambda := \prod_{\substack{q \text{ prime}, \\ q-1 \mid \lambda}} q.$$

For example, $H_{36} = 1919190 = 2 \cdot 3 \cdot 5 \cdot 7 \cdot 13 \cdot 19 \cdot 37$ and $H_{37} = 2$. Note that $H_\lambda$ is always squarefree and $(\lambda+1)$-smooth. We are interested in finding $\lambda$ for which $H_\lambda$ is unusually large; that is, for which $\lambda$ has many divisors $d$ such that $d+1$ happens to be prime. Most of the heavy lifting is done by the following remarkable result.

**LEMMA 4.2.** *Define $\lambda_0(k) := \min \{\lambda \in \mathbb{N}: H_\lambda \geqslant \sqrt{k}\}$. There exist computable absolute constants $c_5 > c_4 > 0$ such that for any integer $k > 100$ we have*

$$(\log k)^{c_4 \log \log \log k} < \lambda_0(k) < (\log k)^{c_5 \log \log \log k}.$$

**Proof.** This is part of [2, Theorem 3]. (The threshold $\sqrt{k}$ could of course be replaced by any fixed power of $k$. It is stated this way in [2] because that paper is concerned with primality testing.) □

The link between $H_\lambda$ and $\mathbb{F}_{p^\lambda}$ is as follows.

**LEMMA 4.3.** *Let $p$ be a prime and let $\lambda \geqslant 2$. Then there exists a $(\lambda+1)$-smooth integer $M \geqslant H_\lambda/(\lambda+1)$ such that $M \mid p^\lambda - 1$.*

**Proof.** We take $M := H_\lambda/p$ if $p$ divides $H_\lambda$, and otherwise $M := H_\lambda$. In the former case we must have $p \leqslant \lambda+1$, so in both cases $M \geqslant H_\lambda/(\lambda+1)$. To see that $M \mid p^\lambda - 1$, consider any prime divisor $q \neq p$ of $H_\lambda$. Then $q-1 \mid \lambda$, so $q \mid p^\lambda - 1$ by Fermat's little theorem. □

**Remark 4.4.** The integer $M$ constructed in Lemma 4.3 only takes into account the structure of $H_\lambda$, and ignores $p^\lambda - 1$ itself. In practice, $p^\lambda - 1$ will often have small factors other than those in $H_\lambda$, possibly including repeated factors (which are never detected by $H_\lambda$). For example, $H_6 = 2 \cdot 3 \cdot 7$, but $19^6 - 1 = 2^3 \cdot 3^3 \cdot 5 \cdot 7^3 \cdot 127$. In an implementation, one should always incorporate these highly valuable "accidental" factors into $M$. We will ignore them in our theoretical discussion.

Next we give a simple sieving algorithm that tabulates approximations of $\log H_\lambda$ for all $\lambda$ up to a prescribed bound.

**LEMMA 4.5.** *Let $m \geqslant 2$. In time $O(m^2)$, we may compute integers $\ell_1, \ldots, \ell_m$ with*

$$|\ell_\lambda - \log H_\lambda| \leqslant 1, \qquad (1 \leqslant \lambda \leqslant m). \tag{4.1}$$

**Proof.** Initialise a table of $\ell_\lambda$ with $\ell_\lambda := 0$ for $\lambda = 1, \ldots, m$. Since $\log H_\lambda \leqslant \sum_{q \leqslant \lambda+1} \log q = O(\lambda)$, it suffices to set aside $O(\lg m)$ bits for each $\ell_\lambda$.

For each integer $q = 2, \ldots, m$, perform the following steps. First test whether $q$ is prime, discarding it if not. Using trial division, the cost is $q^{1/2+o(1)}$, so $m^{3/2+o(1)} = O(m^2)$ overall. Now assume that $q$ is found to be prime. Using a fast algorithm for computing logarithms [8], compute an integer $L_q$ such that $|L_q - 2^r \log q| \leqslant 1$, where $r := 1 + \lg m$, in time $O((\lg m)^{1+o(1)})$. Scan through the table, replacing $\ell_\lambda$ by $\ell_\lambda + L_q$ for those $\lambda$ divisible by $q-1$, i.e., every $(q-1)$-th entry, in time $O(m \lg m)$. The total cost for each prime $q$ is $O(m \lg m)$, and there are $O(m/\lg m)$ primes, so the overall cost is $O(m^2)$. At the end, we divide each $\ell_\lambda$ by $2^r$, yielding the required approximations satisfying (4.1). □

**Proof of Theorem 4.1.** We are given $p$ and $n$ as input. Applying Lemma 4.2 with $k := n^3$, we find that for large enough $n$ there exists some $\lambda$ with

$$(\log (n^3))^{c_4 \log \log \log (n^3)} < \lambda < (\log (n^3))^{c_5 \log \log \log (n^3)}$$

and such that $H_\lambda \geqslant n^{3/2}$. Choose any positive $c_2 < c_4 \log 2$ and any $c_3 > c_5 \log 2$. Since

$$\log\log(n^3) \log\log\log(n^3) \;\sim\; (\log 2) \log \lg n \lg \lg \lg n,$$

we have

$$(\lg n)^{c_2 \lg \lg \lg n} < (\log(n^3))^{c_4 \log \log \log(n^3)} < \lambda < (\log(n^3))^{c_5 \log \log \log(n^3)} < (\lg n)^{c_3 \lg \lg \lg n}$$

for large $n$. Using Lemma 4.5, we may find one such $\lambda$ in time $O(m^2)$, where $m := (\lg n)^{c_3 \lg \lg \lg n}$.

Let $M$ be as in Lemma 4.3. Then $M$ divides $p^\lambda - 1$, and

$$M \geqslant \frac{H_\lambda}{\lambda+1} \geqslant \frac{n^{3/2}}{m+1} \geqslant n$$

for large $n$. We may compute the prime factorisation of $M$ by simply enumerating the primes $q \leqslant \lambda + 1$, $q \neq p$, and checking whether $q - 1$ divides $\lambda$ for each $q$. This can be done in time $O(\lambda^2) = O(m^2)$. □

The main multiplication algorithm depends on a reduction of a "long" DFT to many "short" DFTs. It is essential to have some control over the long and short transform lengths. The following result packages together the prime divisors of the abovementioned $M$, to obtain a long transform length $N$ near a given target $L$, and short transform lengths $N_i$ near a given target $S$.

THEOREM 4.6. *Let $p$, $n$, $\lambda$, $M$ be as in Theorem 4.1. Let $L$ and $S$ be positive integers such that $\lambda < S < L < M$. Then there exist $(\lambda+1)$-smooth integers $N_1, \ldots, N_d$ with the following properties:*

a) $N := N_1 \cdots N_d$ divides $M$ (and hence divides $p^\lambda - 1$).

b) $L \leqslant N \leqslant (\lambda+1) L$.

c) $S \leqslant N_i \leqslant S^3$ for all $i$.

*Given $\lambda$, $S$, $L$, and the prime factorisation of $M$, we may compute such $N_1, \ldots, N_d$ (and their factorisations) in time $\tilde{O}(\lambda^3)$.*

**Proof.** Let $M = N_1 \cdots N_s$ be the prime decomposition of $M$. Taking $d$ minimal with $N_1 \cdots N_d \geqslant L$, we ensure that (a) and (b) are satisfied. At this stage we have $N_i \leqslant \lambda + 1 \leqslant S$ for all $i$. As long as the tuple $(N_1, \ldots, N_d)$ contains an entry $N_i$ with $N_i < S$, we pick the two smallest entries $N_i$ and $N_j$ and replace them by a single entry $N_i N_j$. Clearly, this does not alter the product $N$ of all entries, so (a) and (b) continue to hold. Furthermore, as long as there exist two entries $N_i$ and $N_j$ with $N_i < S$ and $N_j < S$, new entries $N_i N_j$ will always be smaller than $S^2$. Only at the very last step of the loop, the second smallest entry $N_j$ might be larger than $S$. In that case, the product of the entry $N_i$ with $N_i < S$ with $N_j$ is still bounded by $S \cdot S^2 = S^3$. This shows that condition (c) is satisfied at the end of the loop.

Determining $d$ requires at most $s = O(\lambda)$ multiplications of integers less than $M$. There are at most $s$ iterations of the replacement loop. Each iteration must scan through at most $s$ integers of bit size $O(\log M) = O(\lambda)$ and perform one multiplication on such integers. □

## 5. LOGARITHMICALLY SLOW RECURRENCE INEQUALITIES

Let $\Phi: (x_0, \infty) \to \mathbb{R}$ be a smooth increasing function, for some $x_0 \in \mathbb{R}$. We say that $\Phi^*: (x_0, \infty) \to \mathbb{R}^{\geqslant}$ is an *iterator* of $\Phi$ if $\Phi^*$ is increasing and if

$$\Phi^*(x) \;=\; \Phi^*(\Phi(x)) + 1 \tag{5.1}$$

for all sufficiently large $x$.

For instance, the standard iterated logarithm $\log^*$ defined in (1.2) is an iterator of $\log$. An analogous iterator may be defined for any smooth increasing function $\Phi: (x_0, \infty) \to \mathbb{R}$ for which there exists some $\sigma \geqslant x_0$ such that $\Phi(x) \leqslant x - 1$ for all $x > \sigma$. Indeed, in that case,

$$\Phi^*(x) \;:=\; \min \{k \in \mathbb{N} : \Phi^{\circ k}(x) \leqslant \sigma\}$$

is well-defined and satisfies (5.1) for all $x > \sigma$. It will sometimes be convenient to increase $x_0$ so that $\Phi(x) \leqslant x - 1$ is satisfied on the whole domain of $\Phi$.

We say that $\Phi$ is *logarithmically slow* if there exists an $\ell \in \mathbb{N}$ such that

$$(\log^{\circ \ell} \circ \, \Phi \circ \exp^{\circ \ell})(x) \;=\; \log x + O(1) \tag{5.2}$$

for $x \to \infty$. For example, the functions $\log(2x)$, $2 \log x$, $(\log x)^2$ and $(\log x)^{\log \log x}$ are logarithmically slow, with $\ell = 0, 1, 2, 3$ respectively.

LEMMA 5.1. *Let $\Phi \colon (x_0, \infty) \to \mathbb{R}$ be a logarithmically slow function. Then there exists $\sigma \geqslant x_0$ such that $\Phi(x) \leqslant x - 1$ for all $x > \sigma$. Consequently all logarithmically slow functions admit iterators.*

In this paper, the main role played by logarithmically slow functions is to measure *size reduction* in multiplication algorithms. In other words, multiplication of objects of size $n$ will be reduced to multiplication of objects of size $n'$, where $n' \leqslant \Phi(n)$ for some logarithmically slow function $\Phi(x)$. The following result asserts that, from the point of view of iterators, such functions are more or less interchangeable with $\log x$.

LEMMA 5.2. *For any iterator $\Phi^*$ of a logarithmically slow function $\Phi$, we have*

$$\Phi^*(x) \;=\; \log^* x + O(1).$$

The next result is our main tool for converting recurrence inequalities into actual asymptotic bounds for solutions.

PROPOSITION 5.3. *Let $K > 1$, $B \geqslant 0$ and $\ell \in \mathbb{N}$. Let $x_0 \geqslant \exp^{\circ \ell}(1)$, and let $\Phi \colon (x_0, \infty) \to \mathbb{R}$ be a logarithmically slow function such that $\Phi(x) \leqslant x - 1$ for all $x > x_0$. Then there exists a positive constant $C$ (depending on $x_0$, $\Phi$, $K$, $B$ and $\ell$) with the following property.*

*Let $\sigma \geqslant x_0$ and $L > 0$. Let $\mathcal{S} \subseteq \mathbb{R}$, and let $T \colon \mathcal{S} \to \mathbb{R}^{\geqslant}$ be any function satisfying the following recurrence. First, $T(y) \leqslant L$ for all $y \in \mathcal{S}$, $y \leqslant \sigma$. Second, for all $y \in \mathcal{S}$, $y > \sigma$, there exist $y_1, \dots, y_d \in \mathcal{S}$ with $y_i \leqslant \Phi(y)$, and weights $\gamma_1, \dots, \gamma_d \geqslant 0$ with $\sum_i \gamma_i = 1$, such that*

$$T(y) \;\leqslant\; K \left( 1 + \frac{B}{\log^{\circ \ell} y} \right) \sum_{i=1}^{d} \gamma_i \, T(y_i) + L.$$

*Then we have $T(y) \leqslant C L K^{\log^* y - \log^* \sigma}$ for all $y \in \mathcal{S}$, $y > \sigma$.*

## 6. THE CRANDALL–FAGIN ALGORITHM

Consider the problem of computing the product of two $n$-bit integers modulo $m := 2^n - 1$. The naive approach is to compute their ordinary $2n$-bit product and then reduce modulo $m$. The reduction cost is negligible because of the special form of $m$. If $n$ is divisible by a high power of two, one can save a factor of two by using the fact that FFTs naturally compute cyclic convolutions.

An ingenious algorithm of Crandall and Fagin [13] allows for the gain of this precious factor of two for *arbitrary $n$*. Their algorithm is routinely used in the extreme case where $n$ is prime, in the large-scale GIMPS search for Mersenne primes (see http://www.mersenne.org/).

A variant of the Crandall–Fagin algorithm was a key ingredient of our algorithm for integer multiplication that conjecturally achieves the bound $\mathsf{I}(n) = O(n \log n \, 4^{\log^* n})$ [18, Section 9]. In this section we present yet another variant, for computing products in $\mathbb{F}_p[X]/(X^n - 1)$.

### 6.1. The Crandall–Fagin reduction for polynomials

Let $n$, $N$ and $\kappa$ be positive integers with $1 \leqslant N \leqslant n$ and $\kappa \geqslant 2 \lceil n/N \rceil$. Our aim is to reduce multiplication in $\mathbb{F}_p[X]/(X^n - 1)$ to multiplication in $\mathbb{F}_{p^\kappa}[Y]/(Y^N - 1)$. In the applications in subsequent sections, $N$ will be a divisor of $p^\kappa - 1$ with many small factors, so that we can multiply efficiently in $\mathbb{F}_{p^\kappa}[Y]/(Y^N - 1)$ using FFTs over $\mathbb{F}_{p^\kappa}$.

Suppose that $\mathbb{F}_{p^\kappa}$ is represented as $\mathbb{F}_p[Z]/P$, for some irreducible $P \in \mathbb{F}_p[Z]$ of degree $\kappa$. The reduction relies on the existence of an element $\theta \in \mathbb{F}_{p^\kappa}$ such that $\theta^N = Z$. (This is the analogue of the real $N$-th root of 2 appearing in the usual Crandall–Fagin algorithm, which was originally formulated over $\mathbb{C}$.) It is easy to see that such $\theta$ may not always exist; for example, there is no cube root of $Z$ in $\mathbb{F}_{16} = \mathbb{F}_2[Z]/(Z^4 + Z + 1)$. Nevertheless, the next result shows that if $\kappa$ is large enough, then we can always find some modulus $P$ for which a suitable $\theta$ exists.

PROPOSITION 6.1. *Suppose that $p^{\kappa/2} > N$. Then we may compute an irreducible polynomial $P \in \mathbb{F}_p[Z]$ of degree $\kappa$, and an element $\theta \in \mathbb{F}_{p^\kappa} = \mathbb{F}_p[Z]/P$ such that $\theta^N = Z$, in time $\tilde{O}(\kappa^9 p)$.*

**Proof.** We first observe that if $\zeta \in \mathbb{F}_{p^\kappa}$ is a primitive root, then $\zeta^N$ cannot lie in a proper subfield of $\mathbb{F}_{p^\kappa}$. (This property is independent of $P$.) Indeed, if $\zeta^N \in \mathbb{F}_{p^\ell}$ for some proper divisor $\ell$ of $\kappa$, then every $N$-th power in $\mathbb{F}_{p^\kappa}$ lies in $\mathbb{F}_{p^\ell}$. This contradicts the fact that the number of $N$-th powers in $\mathbb{F}_{p^\kappa}$ is at least $1 + (p^\kappa - 1)/N \geqslant p^\kappa/N > p^{\kappa/2} \geqslant p^\ell$.

Now we give an algorithm for computing $P$ and $\theta$. We start by using Lemma 3.1 to compute an irreducible $\tilde{P} \in \mathbb{F}_p[U]$ of degree $\kappa$ in time $\tilde{O}(\kappa^4 p^{1/2})$, and we temporarily represent $\mathbb{F}_{p^\kappa}$ as $\mathbb{F}_p[U]/\tilde{P}$.

Using Lemma 3.2, we may construct a list of $\tilde{O}(\kappa^6 p)$ candidates for primitive roots of $\mathbb{F}_{p^\kappa}$ (see the proof of Lemma 3.3). For each candidate $\zeta$, we compute $\zeta^N$ and its powers $1, \zeta^N, ..., \zeta^{(\kappa-1)N}$. If these are linearly dependent over $\mathbb{F}_p$ then $\zeta^N$ belongs to a proper subfield (and so, as shown above, $\zeta$ is not actually a primitive root). Thus we must eventually encounter some $\zeta$ for which they are linearly independent. We take $P$ to be the minimal polynomial of this $\zeta^N$, which may be computed by finding a linear relation among $1, \zeta^N, ..., \zeta^{\kappa N}$. Let $\varphi: \mathbb{F}_p[Z]/P \to \mathbb{F}_p[U]/\tilde{P}$ be the isomorphism that sends $Z$ to $\zeta^N$. The matrix of $\varphi$ with respect to the standard bases $1, Z, ..., Z^{\kappa-1}$ and $1, U, ..., U^{\kappa-1}$ is given by the coefficients of $1, \zeta^N, ..., \zeta^{(\kappa-1)N}$. The inverse of this matrix yields the matrix of $\varphi^{-1}: \mathbb{F}_p[U]/\tilde{P} \to \mathbb{F}_p[Z]/P$. We then set $\theta = \varphi^{-1}(\zeta)$, so that $\theta^N = \varphi^{-1}(\zeta^N) = Z$.

For each candidate $\zeta$, the cost of computing the necessary powers of $\zeta$ is $\tilde{O}((\kappa + \lg N) \kappa \lg p) = \tilde{O}(\kappa^2 \lg^2 p)$, and the various linear algebra steps require time $\tilde{O}(\kappa^3 \lg p)$ using classical matrix arithmetic. □

In the remainder of this section, we fix some $P$ and $\theta$ as in Proposition 6.1, and assume that $\mathbb{F}_{p^\kappa}$ is represented as $\mathbb{F}_p[Z]/P$. Suppose that we wish to compute the product of $u, v \in \mathbb{F}_p[X]/(X^n - 1)$. The presentation here closely follows that of [18, Section 9.2]. We decompose $u$ and $v$ as

$$u = \sum_{i=0}^{N-1} u_i X^{e_i}, \qquad v = \sum_{i=0}^{N-1} v_i X^{e_i}, \qquad (6.1)$$

where

$$e_i := \lceil n i / N \rceil,$$
$$u_i, v_i \in \mathbb{F}_p[X]_{e_{i+1} - e_i}.$$

Notice that $e_{i+1} - e_i$ takes only two possible values: $\lfloor n/N \rfloor$ or $\lceil n/N \rceil$.

For $0 \leqslant i < N$, let

$$c_i := N e_i - n i, \qquad (6.2)$$

so that $0 \leqslant c_i < N$. For any $0 \leqslant i_1, i_2 < N$, define $\delta_{i_1, i_2} \in \mathbb{Z}$ as follows. Choose $\sigma \in \{0, 1\}$ so that $i := i_1 + i_2 - \sigma N$ lies in the interval $0 \leqslant i < N$, and put

$$\delta_{i_1, i_2} := e_{i_1} + e_{i_2} - e_i - \sigma n.$$

From (6.2), we have

$$c_{i_1} + c_{i_2} - c_i = N(e_{i_1} + e_{i_2} - e_i) - n(i_1 + i_2 - i) = N \delta_{i_1, i_2}.$$

Since the left hand side lies in the interval $(-N, 2N)$, this shows that $\delta_{i_1, i_2} \in \{0, 1\}$. Now, since $e_{i_1} + e_{i_2} \equiv e_i + \delta_{i_1, i_2} \pmod{n}$, we have

$$u v = \sum_{i_1=0}^{N-1} \sum_{i_2=0}^{N-1} u_{i_1} v_{i_2} X^{e_{i_1} + e_{i_2}} = \sum_{i=0}^{N-1} w_i X^{e_i} \pmod{X^n - 1},$$

where
$$w_i := \sum_{i_1+i_2 \equiv i \,(\mathrm{mod}\, N)} X^{\delta_{i_1,i_2}} u_{i_1} v_{i_2}.$$

Since $u_{i_1} \in \mathbb{F}_p[X]_{\lceil n/N \rceil}$ and similarly for $v_{i_2}$, we have $w_i \in \mathbb{F}_p[X]_{2\lceil n/N \rceil}$. Note that we may recover $uv$ from $w_0, ..., w_{N-1}$ in time $O(n \lg p)$, by a standard overlap-add procedure (provided that $N = O(n/\lg n)$).

Now, regarding $u_i$ and $v_i$ as elements of $\mathbb{F}_{p^\kappa}$ by sending $X$ to $Z$, define polynomials $U, V \in \mathbb{F}_{p^\kappa}[Y]/(Y^N - 1)$ by $U_i := \theta^{c_i} u_i$ and $V_i := \theta^{c_i} v_i$ for $0 \leqslant i < N$, and let $W = W_0 + \cdots + W_{N-1} Y^{N-1} := UV$ be their (cyclic) product. Then
$$\tilde{w}_i := \theta^{-c_i} W_i = \sum_{i_1+i_2 \equiv i \,(\mathrm{mod}\, N)} \theta^{-c_i} U_{i_1} V_{i_2} = \sum \theta^{c_{i_1}+c_{i_2}-c_i} u_{i_1} v_{i_2} = \sum Z^{\delta_{i_1,i_2}} u_{i_1} v_{i_2}$$

coincides with the reinterpretation of $w_i$ as an element of $\mathbb{F}_{p^\kappa}$. Moreover, we may recover $w_i$ unambiguously from $\tilde{w}_i$, as $\kappa \geqslant 2 \lceil n/N \rceil$ and $w_i \in \mathbb{F}_p[X]_{2\lceil n/N \rceil}$. Altogether, this shows how to reduce multiplication in $\mathbb{F}_p[X]/(X^n - 1)$ to multiplication in $\mathbb{F}_{p^\kappa}[Y]/(Y^N - 1)$.

**Remark 6.2.** The pair $(e_{i+1}, c_{i+1})$ can be computed from $(e_i, c_i)$ in $O(\lg n)$ bit operations, so we may compute the sequences $e_0, ..., e_{N-1}$ and $c_0, ..., c_{N-1}$ in time $O(N \lg n)$. Moreover, since $c_{i+1} - c_i$ takes on only two possible values, we may compute the sequence $\theta^{c_0}, ..., \theta^{c_{N-1}}$ using $O(N)$ multiplications in $\mathbb{F}_{p^\kappa}$.

## 7. The main algorithm

Consider the problem of computing $t \geqslant 1$ products $u_1 v, ..., u_t v$ with $u_1, ..., u_t, v \in \mathbb{F}_p[X]/(X^n - 1)$, i.e., $t$ products with one fixed operand. Denote the cost of this operation by $\mathsf{C}_{p,t}(n)$. Our algorithm for this problem will perform $t+1$ forward DFTs and $t$ inverse DFTs, so it is convenient to introduce the normalisation
$$\mathsf{C}_p(n) := \sup_{t \geqslant 1} \frac{\mathsf{C}_{p,t}(n)}{2t+1}.$$

This is well-defined since clearly $\mathsf{C}_{p,t}(n) \leqslant t\, \mathsf{C}_{p,1}(n)$. Roughly speaking, $\mathsf{C}_p(n)$ may be thought of as the notional cost of a single DFT.

The problem of multiplying two polynomials in $\mathbb{F}_p[X]$ of degree less than $k$ may be reduced to the above problem by using zero-padding, i.e., taking $n := 2k$ and $t := 1$. Since $\mathsf{C}_{p,1}(n) \leqslant 3\, \mathsf{C}_p(n)$, we obtain $\mathsf{M}_p(k) \leqslant 3\, \mathsf{C}(2k) + O(k \lg p)$. Thus it suffices to obtain a good bound for $\mathsf{C}_p(n)$.

The next theorem gives the core of the new algorithm for the case that $n$ is large relative to $p$.

**Theorem 7.1.** *There exist $x_0 \geqslant 2$ and a logarithmically slow function $\Phi: (x_0, \infty) \to \mathbb{R}$ with the following property. For all integers $n > x_0$, there exist integers $n_1, ..., n_d \leqslant \Phi(n)$, and weights $\gamma_1, ..., \gamma_d \geqslant 0$ with $\sum_i \gamma_i = 1$, such that*
$$\frac{\mathsf{C}_p(n)}{n \lg p \lg (n \lg p)} \leqslant \left(8 + O\left(\frac{1}{\lg \lg n}\right)\right) \sum_{i=1}^d \gamma_i \frac{\mathsf{C}_p(n_i)}{n_i \lg p \lg(n_i \lg p)} + O(1), \quad (7.1)$$

*uniformly for $n > \max(x_0, p^2)$.*

**Proof.** We wish to compute $t \geqslant 1$ products $u_1 v, ..., u_t v$ with $u_1, ..., u_t, v \in \mathbb{F}_p[X]/(X^n - 1)$, for some sufficiently large $n$. We assume throughout that $p^2 < n$.

**Choose parameters.** Using Theorem 4.1, we obtain integers $\lambda$ and $M$ with
$$(\lg n)^{c_2 \lg \lg \lg n} < \lambda < (\lg n)^{c_3 \lg \lg \lg n},$$

and so that $M \geqslant n$, $M \,|\, p^\lambda - 1$, and $M$ is $(\lambda+1)$-smooth. We choose long and short target transform lengths $L := \lceil n/\lambda^3 \rceil$ and $S := \lambda^{(\lg \lg n)^2}$. For large enough $n$ we then have $\lambda < S < L < n \leqslant M$, so we may apply Theorem 4.6. This yields $(\lambda+1)$-smooth integers $N_1, ..., N_d$, with known factorisations, such that $N := N_1 \cdots N_d$ divides $p^\lambda - 1$ and lies in the range $L \leqslant N \leqslant (\lambda + 1)L$, and such that $S \leqslant N_i \leqslant S^3$ for all $i$. Finally we set $\kappa := \lceil 2 \lceil n/N \rceil / \lambda \rceil \lambda$, so that $N$ also divides $p^\kappa - 1$, and we put $n_i := 2 N_i \kappa$. All of these parameters may be computed in time $O((\lg n)^{\lg \lg n}) = O(n)$.

For large $n$, we observe that the following estimates hold. First, we have
$$\frac{n}{\lambda^3} \leqslant N \leqslant \frac{2n}{\lambda^2}.$$

Since $\log \lambda \asymp \lg \lg n \lg \lg \lg n$, it follows that
$$\log_2 N = \left(1 + O\!\left(\frac{1}{\lg \lg n}\right)\right) \lg n,$$

and also that $\log \kappa \asymp \lg \lg n \lg \lg \lg n$. We have $N\kappa = 2n + O(N\lambda) = 2n + O(n/\lambda)$, so
$$N\kappa = \left(2 + O\!\left(\frac{1}{\lg n}\right)\right) n.$$

To estimate $d$, note that $N \geqslant S^d$ and $\log S \asymp (\lg \lg n)^3 \lg \lg \lg n$, so
$$d = O\!\left(\frac{\lg n}{(\lg \lg n)^3 \lg \lg \lg n}\right).$$

Since $p \leqslant n$ we have
$$\lg(n \lg p) = \left(1 + O\!\left(\frac{1}{\lg \lg n}\right)\right) \lg n, \tag{7.2}$$

and as $n_i \geqslant N_i \geqslant S \geqslant (\lg n)^{\lg \lg n}$, we similarly obtain
$$\lg(n_i \lg p) = \left(1 + O\!\left(\frac{1}{\lg \lg n}\right)\right) \lg n_i. \tag{7.3}$$

**Crandall–Fagin reduction.** Let us check that the hypotheses of section 6.1 are satisfied, to enable the reduction to multiplication in $\mathbb{F}_{p^\kappa}[Y]/(Y^N - 1)$. We certainly have $1 \leqslant N \leqslant n$ and $\kappa \geqslant 2\lceil n/N \rceil$. For Proposition 6.1, observe that $\kappa \geqslant \lambda^2 \succ \lg n \sim \lg N$, so $p^{\kappa/2} \geqslant 2^{\kappa/2} > N$ for large $n$. Thus we obtain an irreducible $P \in \mathbb{F}_p[Z]$ of degree $\kappa$, and an element $\theta \in \mathbb{F}_{p^\kappa} = \mathbb{F}_p[Z]/P$ such that $\theta^N = Z$, in time $\tilde{O}(\kappa^9 p) = \tilde{O}(\lambda^{27} n^{1/2}) = O(n)$. Each multiplication in $\mathbb{F}_{p^\kappa}$ has cost $O(\mathsf{S}_p(\kappa))$ (see section 3).

Computing the sequences $e_i$ and $c_i$ costs $O(N \lg n) = O(n \lg n)$, and computing the sequence $\theta^{c_i}$ costs $O(N \mathsf{S}_p(\kappa))$. The initial splitting and final overlap-add phases require time $O(t n \lg p)$, and the multiplications by $\theta^{c_i}$ and $\theta^{-c_i}$ cost $O(t N \mathsf{S}_p(\kappa))$.

**Long transforms.** The factorisation of $N$ is known, and $N$ divides $p^\kappa - 1$, so by Lemma 3.3 we may compute a primitive $N$-th root of unity $\omega \in \mathbb{F}_{p^\kappa}$ in time $\tilde{O}(\kappa^9 p) = O(n)$.

We will multiply in $\mathbb{F}_{p^\kappa}[Y]/(Y^N - 1)$ by using DFTs with respect to $\omega$. The table of roots $1, \omega, \ldots, \omega^{N-1}$ may be computed in time $O(N \mathsf{S}_p(\kappa))$. In a moment, we will describe an algorithm $\mathcal{A}_i$ for computing a "short" DFT of length $N_i$ with respect to $\omega_i := \omega^{N/N_i}$; we then use the algorithm $\mathcal{A} := \mathcal{A}_1 \odot \cdots \odot \mathcal{A}_d$ for the main transform of length $N$ (see section 2.3). The corresponding twiddle factor tables may be extracted in time $O(N \lg N (\kappa \lg p + \lg N)) = O(n \lg n \lg p)$.

Let $\mathsf{D}$ denote the complexity of $\mathcal{A}$, and for $t' \geqslant 1$ let $\mathsf{D}_{i,t'}$ denote the cost of performing $t'$ independent DFTs of length $N_i$ using $\mathcal{A}_i$. Then by (2.4) we have
$$\mathsf{D} \leqslant \sum_{i=1}^d \mathsf{D}_{i, N/N_i} + O(d N \mathsf{S}_p(\kappa)) + O((\kappa \lg p) N \lg N).$$

The last term is simply $O(n \lg n \lg p)$.

**Bluestein conversion.** We now begin constructing $\mathcal{A}_i$, assuming that $t' \geqslant 1$ independent transforms are sought. We first use Bluestein's algorithm (section 2.5) to convert each DFT of length $N_i$ to a multiplication in $\mathbb{F}_{p^\kappa}[X]/(X^{N_i} - 1)$. We must check that 2 is invertible in $\mathbb{F}_{p^\kappa}$ if $N_i$ is even; indeed, if $N_i$ is even, then so is $p^\kappa - 1$, so $p \neq 2$. The Bluestein conversion contributes $O(t' N_i \mathsf{S}_p(\kappa))$ to the cost of $\mathcal{A}_i$.

We must also compute a suitable table of roots, once at the top level. We first extract the table $1, \omega_i, \ldots, \omega_i^{N_i - 1}$ from the top level table in time $O(N \kappa \lg p) = O(n \lg p)$, and then sort them into the correct order (and perform any necessary additions) in time $O(N_i \lg N_i (\kappa \lg p + \lg N)) = O(S^3 \lg N_i \kappa \lg p) = O(\lg N_i (n \lg p))$. Over all $i$ the cost is $O(\lg N (n \lg p)) = O(n \lg n \lg p)$.

**Kronecker substitution.** We finally convert each multiplication in $\mathbb{F}_{p^\kappa}[X]/(X^{N_i}-1)$ to a multiplication in $\mathbb{F}_p[X]/(X^{n_i}-1)$ using Kronecker substitution (see section 2.6). The latter multiplications have cost $\mathsf{C}_{p,t'}(n_i)$, since one argument is fixed. After the multiplications, we must also perform $t'N_i$ divisions by $P$ to recover the results in $\mathbb{F}_{p^\kappa}[X]/(X^{N_i}-1)$, at cost $O(t'N_i\mathsf{S}_p(\kappa))$. Consolidating the estimates for the Bluestein conversion and Kronecker substitution, we have

$$\begin{aligned}\mathsf{D}_{i,t'} &\leqslant \mathsf{C}_{p,t'}(n_i)+O(t'N_i\mathsf{S}_p(\kappa)).\\ &\leqslant (2\,t'+1)\,\mathsf{C}_p(n_i)+O(t'N_i\mathsf{S}_p(\kappa)).\end{aligned}$$

For $t'=N/N_i\geqslant N/S^3\succ\lg\lg n$, this becomes

$$\mathsf{D}_{i,N/N_i}\leqslant (2+O(1/\lg\lg n))\,(N/N_i)\,\mathsf{C}_p(n_i)+O(N\mathsf{S}_p(\kappa)).$$

**Conclusion.** Summing over $i$ yields

$$\mathsf{D}\leqslant \left(2+O\!\left(\frac{1}{\lg\lg n}\right)\right)\sum_{i=1}^d\frac{N}{N_i}\mathsf{C}_p(n_i)+O(d\,N\,\mathsf{S}_p(\kappa))+O(n\lg n\lg p).$$

Since

$$\begin{aligned}d\,N\,\mathsf{S}_p(\kappa) &= O\!\left(\frac{\lg n}{(\lg\lg n)^3\lg\lg\lg n}N\kappa\lg\kappa\lg\lg\kappa\lg p\lg\lg p\lg\lg\lg p\right)\\ &= O\!\left(n\lg n\lg p\,\frac{(\lg\lg\lg n)^2}{\lg\lg n}\right)\\ &= O(n\lg n\lg p)\end{aligned}$$

and

$$\frac{N}{N_i}=\frac{2\,N\kappa}{2\,N_i\kappa}=\left(4+O\!\left(\frac{1}{\lg n}\right)\right)\frac{n}{n_i},$$

this becomes

$$\mathsf{D}\leqslant \left(8+O\!\left(\frac{1}{\lg\lg n}\right)\right)\sum_{i=1}^d\frac{n}{n_i}\mathsf{C}_p(n_i)+O(n\lg n\lg p).$$

Let $\gamma_i:=\log N_i/\log N$, so that $\sum_i\gamma_i=1$. Since

$$\lg n_i=\log_2 N_i+O(\log\kappa)=\left(1+O\!\left(\frac{\log\lambda}{\log S}\right)\right)\log_2 N_i=\left(1+O\!\left(\frac{1}{\lg\lg n}\right)\right)\log_2 N_i$$

and $\lg n=(1+O(1/\lg\lg n))\log_2 N$, we get

$$\mathsf{D}\leqslant \left(8+O\!\left(\frac{1}{\lg\lg n}\right)\right)\sum_{i=1}^d\gamma_i\frac{n\lg n}{n_i\lg n_i}\mathsf{C}_p(n_i)+O(n\lg n\lg p).$$

To compute the desired $t$ products, we must execute $t+1$ forward transforms, and $t$ inverse transforms. Each product also requires $O(N)$ pointwise multiplications in $\mathbb{F}_{p^\kappa}$ and $O(N)$ multiplications by $1/N$. These have cost $O(N\mathsf{S}_p(\kappa))$, which is absorbed by the $O(n\lg n\lg p)$ term. Thus we obtain

$$\mathsf{C}_{p,t}(n)\leqslant (2\,t+1)\,\mathsf{D}+O(t\,n\lg n\lg p).$$

Dividing by $(2\,t+1)\,n\lg n\lg p$, taking suprema over $t\geqslant 1$, and using (7.2) and (7.3), yields the bound (7.1).

Finally, since

$$n_i=O(S^3\lambda^3)=O\bigl((\lg n)^{3c_3\lg\lg\lg n((\lg\lg n)^2+1)}\bigr)=O\bigl((\lg n)^{(\lg\lg n)^3}\bigr),$$

we have $\log n_i=O((\log\log n)^4)$ and hence $\log\log\log n_i\leqslant \log\log\log\log n+C$ for some constant $C$ and all large $n$. We may then take $\Phi(x):=\exp^{\circ 3}(\log^{\circ 4}x+C)$. $\square$

Now we may prove the main theorem announced in the introduction.

**Proof of Theorem 1.1.** For $n \geqslant 2$ define

$$T_p(n) := \frac{\mathsf{C}_p(n)}{n \lg p \lg(n \lg p)}.$$

It suffices to prove that $T_p(n) = O(8^{\log^*(n \lg p)})$, uniformly in $n$ and $p$.

Let $x_0$ and $\Phi(x)$ be as in Theorem 7.1. Increasing $x_0$ if necessary, we may assume that $\Phi(x) \leqslant x - 1$ for $x > x_0$, and that $x_0 \geqslant \exp(\exp(1))$.

Let $\sigma_p := \max(x_0, p^2)$ for each $p$. We will consider the regions $n \leqslant \sigma_p$ and $n > \sigma_p$ separately. First consider the case $n \leqslant \sigma_p$. There are only finitely many primes $p < (x_0)^{1/2}$, so we may assume that $p^2 \geqslant x_0$ and that $n \leqslant \sigma_p = p^2$. In this region we use Kronecker substitution: by (2.5) and (1.1) we have

$$\mathsf{M}_p(n) = O(\mathsf{I}(n \lg p)) = O\big(n \lg p \lg (n \lg p) \, 8^{\log^*(n \lg p)}\big),$$

and since

$$\mathsf{C}_p(n) = \sup_{t \geqslant 1} \frac{\mathsf{C}_{p,t}(n)}{2t+1} \leqslant \sup_{t \geqslant 1} \frac{t \, \mathsf{C}_{p,1}(n)}{2t+1} = O(\mathsf{C}_{p,1}(n)) = O(\mathsf{M}_p(n)),$$

we get $T_p(n) = O(8^{\log^*(n \lg p)})$ uniformly for $n \leqslant \sigma_p$. In fact, this even shows that $T_p(n) = O(8^{\log^* p})$ uniformly for $n \leqslant \sigma_p$.

Now consider the case $n > \sigma_p$. Here we invoke Theorem 7.1 to obtain absolute constants $B$, $L > 0$ such that for every $n > \sigma_p$, there exist $n_1, ..., n_d \leqslant \Phi(n)$ and $\gamma_1, ..., \gamma_d$ such that

$$T_p(n) \leqslant 8 \left(1 + \frac{B}{\log \log n}\right) \sum_{i=1}^{d} \gamma_i T_p(n_i) + L.$$

Set $L_p := \max(L, \max_{2 \leqslant n \leqslant \sigma_p} T_p(n))$. Applying Proposition 5.3 with $\ell := 2$, $K := 8$, $\mathcal{S} := \{2, 3, ...\}$, we find that $T_p(n) = O(L_p 8^{\log^* n - \log^* \sigma_p})$ uniformly for $n > \sigma_p$. Since $\log^* \sigma_p = \log^* p + O(1)$ and $L_p = O(8^{\log^* p})$, we conclude that $T_p(n) = O(8^{\log^* n}) = O(8^{\log^*(n \lg p)})$ uniformly for $n > \sigma_p$. □

## 8. Conjecturally faster multiplication

Recall that in [18] we established the bound $\mathsf{I}(n) = O(n \lg n \, K^{\log^* n})$ with $K = 8$ for the complexity of integer multiplication. We also proved that this can be improved to $K = 4$, assuming the following slight weakening of the Lenstra–Pomerance–Wagstaff conjecture on the distribution of Mersenne primes [28, 35] (see [18, Section 9] for further discussion).

CONJECTURE 8.1. *Let $\pi_m(x)$ be the number of Mersenne primes less than $x$. Then there exist constants $0 < a < b$ such that for all $x > 3$,*

$$a \log \log x < \pi_m(x) < b \log \log x.$$

The source of the (conjectural) speedup is as follows. The $K = 8$ algorithm of [18, Section 6] computes DFTs over $\mathbb{C}$, and so we encounter the "short product obstruction". Namely, to compute the product of two real numbers with $p$ significant bits using FFT algorithms, we are forced to compute a full product of two $p$-bit integers, and then truncate the result to $p$ bits. To achieve $K = 4$, we replaced the coefficient ring $\mathbb{C}$ by the finite field $\mathbb{F}_p[i]$, where $p = 2^q - 1$ is a Mersenne prime. Multiplication modulo $2^q - 1$ is a "cyclic product" rather than a short product, and this saves a factor of two at each recursion level.

The aim of this section is to outline a credible strategy for achieving the same improvement, from $K = 8$ to $K = 4$, in the context of multiplication in $\mathbb{F}_p[X]$, at least under certain plausible number-theoretic hypotheses.

Consider the problem of multiplying polynomials in $\mathbb{F}_p[X]$ of degree less than $n$. In the Kronecker substitution region, i.e., for $\lg n = O(\lg p)$, we can of course achieve $K = 4$ if we assume Conjecture 8.1.

Now consider the region where $n$ is much larger than $p$. In the algorithm of section 7, we reduced the multiplication problem to computing products in $\mathbb{F}_{p^\kappa}[Y]/(Y^{N_i} - 1)$, where $\mathbb{F}_{p^\kappa} = \mathbb{F}_p[Z]/P$ for some monic irreducible $P \in \mathbb{F}_p[Z]$ of degree $\kappa$, and where $N_i$ is a "short" transform length. These multiplications were in turn converted to multiplications in $\mathbb{F}_p[X]/(X^{2\kappa N_i} - 1)$ via Kronecker substitution. It is exactly this factor of two in zero-padding that we wish to avoid.

Taking our cue from the integer case, we observe that if the modulus $P$ is of a particularly special form, then this factor of two can be eliminated. Indeed, suppose that

$$P(Z) = \frac{Z^{\kappa+1} - 1}{Z - 1} = Z^\kappa + \cdots + Z + 1.$$

Then $\mathbb{F}_{p^\kappa}[Y]/(Y^{N_i} - 1)$ is isomorphic to $\mathbb{F}_p[Y, Z]/(Y^{N_i} - 1, Z^\kappa + \cdots + 1)$, which is a quotient of $\mathbb{F}_p[Y, Z]/(Y^{N_i} - 1, Z^{\kappa+1} - 1)$. Assuming further that $\gcd(N_i, \kappa + 1) = 1$, the latter ring is isomorphic to $\mathbb{F}_p[X]/(X^{(\kappa+1)N_i} - 1)$. In other words, we can reduce a cyclic convolution over $\mathbb{F}_{p^\kappa}$ of length $N_i$ to a cyclic convolution over $\mathbb{F}_p$ of length $(\kappa+1) N_i \sim \kappa N_i$, rather than length $2\kappa N_i$. This saves a factor of two at this recursion level; if we can manage something similar at every recursion level, we reduce $K$ from 8 to 4.

The essential question is therefore how to choose $\kappa$ so that $P = Z^\kappa + \cdots + Z + 1$ is irreducible over $\mathbb{F}_p$. The following lemma gives a simple number-theoretic characterisation of such $\kappa$.

LEMMA 8.2. *Let $\kappa \geqslant 2$. The polynomial $P = Z^\kappa + \cdots + Z + 1$ is irreducible over $\mathbb{F}_p$ if and only if $\kappa + 1$ is prime and $p$ is a primitive root modulo $\kappa + 1$.*

**Proof.** If $\ell$ is a nontrivial factor of $\kappa + 1$, then $Z^{\kappa+1} - 1$ is divisible by $Z^\ell - 1$, so $P$ has the nontrivial factor $Z^{\ell-1} + \cdots + Z + 1$. On the other hand, suppose that $\kappa + 1$ is prime. If $\kappa + 1 = p$, then $P = (Z^p - 1)/(Z - 1) = (Z - 1)^{p-1}$ is not irreducible; otherwise, the number of irreducible factors of $P$ over $\mathbb{F}_p$ is exactly $\kappa/m$, where $m$ is the order of $p$ in $(\mathbb{Z}/(\kappa+1)\mathbb{Z})^*$ [36, Theorem 2.13 and Proposition 2.14] (this last statement is sometimes called the "cyclotomic reciprocity law"). □

A first attempt to reach $K = 4$ might run as follows. First choose $\lambda$ as in the proof of Theorem 7.1, so that $\mathbb{F}_{p^\lambda}$ has many roots of unity of smooth order. Then use the same multiplication algorithm as before, but now working over $\mathbb{F}_{p^\kappa} = \mathbb{F}_p[Z]/(Z^\kappa + \cdots + 1)$, where $\kappa$ is the smallest positive multiple of $\lambda$ such that $q := \kappa + 1$ is prime and such that $p$ is a primitive root modulo $q$.

Unfortunately this plan does not quite work, for reasons that we now explain.

First, if $p \mid \lambda$, there may be a trivial obstruction to the existence of suitable $\kappa$. For example, if $p = 5$ and $5 \mid \lambda$, then for any $q \equiv 1 \pmod{\lambda}$ we have $(5/q) = (q/5) = 1$ by quadratic reciprocity, so 5 cannot be a primitive root modulo $q$. The only way to avoid this seems to be to insist that $\lambda$ not be divisible by $p$ in the first place. Therefore we propose the following strengthening of Lemma 4.2.

CONJECTURE 8.3. *Define*

$$\lambda_0(k, p) := \begin{cases} \min\{\lambda \in \mathbb{N} : X_\lambda \geqslant \sqrt{k}, p \nmid \lambda\} & p \text{ odd,} \\ \min\{\lambda \in \mathbb{N} : X_\lambda \geqslant \sqrt{k}, 8 \nmid \lambda\} & p = 2. \end{cases}$$

*There exist computable absolute constants $c_5' > c_4' > 0$ such that for any integer $k > 100$ and any prime $p$ we have*

$$(\log k)^{c_4' \log \log \log k} < \lambda_0(k, p) < (\log k)^{c_5' \log \log \log k}.$$

It seems likely that this conjecture should be accessible to specialists in analytic number theory. Experimentally, the "missing" factor $p$ does not seem to have much effect on the propensity of $H_\lambda$ to have many divisors $d$ such that $d + 1$ is prime. We could not find a quick way to prove the conjecture directly from Lemma 4.2.

The second problem is more serious. A special case of *Artin's conjecture on primitive roots* (see [25] for a survey) asserts that for any prime $p$, there are infinitely many primes $q$ for which $p$ is a primitive root modulo $q$. Unfortunately, there is not even a single prime $p$ for which Artin's conjecture is known to hold! Therefore, in general we cannot prove existence of a suitable $\kappa$.

However, we do have the following result of Lenstra, which states that the only obstruction to the existence of $\kappa$ is the trivial one noted above, provided we are willing to accept GRH.

LEMMA 8.4. *Assume GRH. Assume that $p \nmid \lambda$ if $p$ is odd, or that $8 \nmid \lambda$ if $p = 2$. Then there exist infinitely many primes $q \equiv 1 \pmod{\lambda}$ such that $p$ is a primitive root modulo $q$.*

**Proof.** This is a special case of [23, Theorem 8.3]. (For a description of exactly which number fields are supposed to satisfy GRH for this result to hold, see [23, Theorem 3.1].) □

The third problem is that existence of $\kappa$ is not enough; we also need a bound on its size, as a function of $\lambda$ and $p$. The literature provides little guidance on this question. Under the hypotheses of Lemma 8.4 (including GRH), it is known that the number of primes $q < x$ such that $q = 1 \pmod{\lambda}$ and for which $p$ is a primitive root modulo $q$ grows as

$$C \frac{x}{\log x} + O\left(\frac{x \log \log x}{\log^2 x}\right) \tag{8.1}$$

for some positive $C$ (see for example [24, Theorem 2]). Both $C$ and the implied big-$O$ constant depend on $\lambda$ and $p$. While the value of $C$ is reasonably well understood, we could not find in the literature a similarly precise description of the implied constant, so we have been unable to derive a bound for $\kappa$.

In the interests of getting our conjectural algorithm off the ground, we make the following guess. Define $q_0(\lambda, p)$ to be the smallest prime $q$ such that $q = 1 \pmod{\lambda}$ and such that $p$ is a primitive root modulo $q$ (assuming that such $q$ exists).

CONJECTURE 8.5. *There exists a constant $\beta \in \mathbb{N}$ and a logarithmically slow function $\Psi(x)$ with the following property. Suppose that $p \nmid \lambda$ if $p$ is odd, or that $8 \nmid \lambda$ if $p = 2$, and suppose that $\log^* \lambda \geqslant \beta + \log^* p$. Then $q_0(\lambda, p) \leqslant \exp \Psi(\lambda)$.*

We have deliberately given an absurdly weak formulation of this conjecture, so as to provide the largest possible target for analytic number theorists. For example, it would be enough to prove that $q_0(\lambda, p) \leqslant \exp((\log \lambda)^{100})$ for all $\lambda \geqslant \exp(\exp(\exp(\exp p)))$. We suspect that the conjecture is accessible under GRH; presumably a proof would require analysing the implied constant in (8.1).

**Remark 8.6.** The reason we need $\lambda$ to be large compared to $p$ is that it is easy to construct, using the Chinese remainder theorem, a prime $p$ which fails to be a primitive root modulo $q$ for all primes $q$ up to a prescribed bound.

**Proof of Theorem 1.2.** The proof is similar to that of Theorem 1.1. We will only give a sketch, highlighting the main differences.

In the region $\lg n = O(\lg p)$, we use Kronecker substitution together with [18, Theorem 2], which states that $\mathsf{I}(n) = O(n \lg n \, 4^{\log^* n})$ under Conjecture 8.1.

Now assume that $\lg p = O(\lg n)$. We will pursue a strategy similar to Theorem 7.1, recursing from $n$ down to an exponentially smaller $\Phi(n)$. Assume that we wish to compute $t \geqslant 1$ products in $\mathbb{F}_p[X]/(X^n - 1)$ with one fixed operand.

Using an appropriate modification of Theorem 4.1, in which the use of Lemma 4.2 is replaced by Conjecture 8.3, we find integers $\lambda$ and $M$ such that $(\lg n)^{c'_4 \lg \lg \lg n} < \lambda < (\lg n)^{c'_5 \lg \lg \lg n}$, $p \nmid \lambda$ (or $8 \nmid \lambda$ if $p = 2$), $M \geqslant n$, $M \mid p^\lambda - 1$, and $M$ is $(\lambda + 1)$-smooth.

Now we wish to apply Conjecture 8.5 to construct suitable $\kappa$. However, that conjecture requires that $\log^* \lambda \geqslant \beta + \log^* p$. If we have instead $\log^* \lambda < \beta + \log^* p$, we simply use the algorithm of Theorem 7.1. This yields the expansion factor $K = 8$, but only for $O(1)$ recursion levels (depending on $\beta$), since $\log^* \lambda = \log^* n + O(1)$. So it is permissible to assume that $\log^* \lambda \geqslant \beta + \log^* p$, losing only a constant factor in the main complexity bound for the last few recursion levels.

Applying Conjecture 8.5 and Lemma 8.2, we obtain a positive multiple $\kappa$ of $\lambda$, such that $\kappa + 1$ is prime, $\kappa < \exp \Psi(\lambda)$, and $P := Z^\kappa + \cdots + 1$ is irreducible in $\mathbb{F}_p[Z]$. We will take $\mathbb{F}_p[Z]/P$ as our model for $\mathbb{F}_{p^\kappa}$; note that $M \mid p^\kappa - 1$.

Notice that $\kappa < \Psi'(n)$ for some logarithmically slow function $\Psi'(x)$. Indeed, there exists $\ell \in \mathbb{N}$ and $C, C' > 0$ such that $\Psi(\lambda) \leqslant \exp^{\circ \ell}(\log^{\circ (\ell+1)}(\lambda) + C)$ and $\log \log \log \lambda \leqslant \log \log \log \log n + C'$ for large $n$. Increasing $\ell$ if necessary, we get $\kappa < \exp \Psi(\lambda) \leqslant \exp^{\circ(\ell+1)}(\log^{\circ(\ell+2)}(n) + C'')$, the latter being a logarithmically slow function of $n$.

In the proof of Theorem 7.1, we selected the transform length $N$ first, and then chose $\kappa$ to fine-tune the total bit size. Here we have less control over $\kappa$, so we must use a different strategy. Choose target long and short transform lengths $L := \lceil n/(\lambda^2 \kappa) \rceil$ and $S := \kappa^{(\lg \lg n)^2 \lg \lg \kappa}$. Applying Theorem 4.6, we obtain $(\lambda + 1)$-smooth integers $N_1, \ldots, N_d$ such that $N := N_1 \cdots N_d$ divides $p^\kappa - 1$ and $S \leqslant N_i \leqslant S^3$ for each $i$, and such that $n/(\lambda^2 \kappa) \leqslant N \leqslant 2n/(\lambda \kappa)$ for large $n$. Let $\mu' := \lceil 2n/(N\kappa) \rceil \geqslant \lambda$, and let $\mu$ be the smallest prime greater than $\mu' + 1$ and different to $\kappa + 1$. By [4] we have $\mu \leqslant \mu' + O((\mu')^{0.525}) \leqslant (1 + O(1/\lg \lg n)) \mu'$, so $\mu N \kappa = (2 + O(1/\lg \lg n)) n$.

We now apply the Crandall–Fagin algorithm to reduce multiplication in $\mathbb{F}_p[X]/(X^n-1)$ to multiplication in $\mathbb{F}_{p^\kappa}[Y]/(Y^{\mu N}-1)$. The prerequisite $\kappa \geqslant 2\lceil n/\mu N\rceil$ is satisfied (since $\kappa$ is even). Instead of using Proposition 6.1 to construct $\theta \in \mathbb{F}_{p^\kappa}$ satisfying $\theta^{\mu N} = Z$, we may simply take $\theta := Z^{(\mu N)^{-1} \bmod (\kappa+1)}$. The required modular inverse exists, as $\mu$ and $\kappa+1$ are distinct primes, and $\gcd(N, \kappa+1) = 1$ since $N$ is $(\lambda+1)$-smooth.

We now take advantage of the isomorphism $\mathbb{F}_{p^\kappa}[Y]/(Y^{\mu N}-1) \cong \mathcal{R}[Y]/(Y^N-1)$, where $\mathcal{R} := \mathbb{F}_{p^\kappa}[U]/(U^\mu - 1)$. We multiply in $\mathcal{R}[Y]/(Y^N-1)$ by using DFTs over $\mathcal{R}$: one DFT of length $N$ over $\mathcal{R}$ reduces to $\mu$ DFTs of length $N$ over $\mathbb{F}_{p^\kappa}$. The latter are handled by decomposing into short transforms of length $N_i$, which are in turn converted to multiplications in $\mathbb{F}_{p^\kappa}[Y]/(Y^{N_i}-1)$ via Bluestein's algorithm. Finally — and this is where all the hard work pays off — each such multiplication reduces to a multiplication in $\mathbb{F}_p[X]/(X^{(\kappa+1)N_i}-1)$, since $\gcd(N_i, \kappa+1) = 1$. The multiplications in $\mathcal{R}$ itself are handled using the algebraic Schönhage–Strassen algorithm.

We omit the rest of the complexity argument, which is essentially the same as that of Theorem 7.1 and Theorem 1.1. We mention only that $S$ was chosen sufficiently large that the multiplications in $\mathcal{R}$ and in $\mathbb{F}_{p^\kappa}$ make a negligible contribution overall. □

## 9. Notes and generalisations

In this section we outline some directions along which the results in this paper can be extended. We also provide some hints concerning the practical usefulness of the new ideas. Our treatment is more sketchy and we plan to provide more details in a forthcoming paper.

### 9.1. Multiplication of polynomials over $\mathbb{F}_{p^\kappa}$

Recall that $\mathsf{M}_{p^\kappa}(n)$ denotes the cost of multiplying polynomials in $\mathbb{F}_{p^\kappa}[X]$ of degree less than $n$, where we assume that some model $\mathbb{F}_p[Z]/P$ for $\mathbb{F}_{p^\kappa}$ has been fixed in advance.

Theorem 9.1. *We have*
$$\mathsf{M}_{p^\kappa}(n) = O\bigl(n\kappa \lg p \lg(n\kappa \lg p)\, 8^{\log^*(n\kappa \lg p)}\bigr),$$
*uniformly for all $n, \kappa \geqslant 1$ and all primes $p$. Assuming Conjectures 8.1, 8.3 and 8.5, we may replace $K=8$ by $K=4$.*

Indeed, we saw in section 3 that
$$\mathsf{M}_{p^\kappa}(n) = \mathsf{M}_p(2n\kappa) + O(n\,\mathsf{D}_p(\kappa)),$$
where $\mathsf{D}_p(\kappa)$ denotes the cost of dividing a polynomial of degree less than $2\kappa$ by $P$. Having established the bound $\mathsf{M}_p(n) = O(n\lg p \lg(n\lg p)\, 8^{\log^*(n\lg p)})$, it is now permissible to assume that $\mathsf{M}_p(n)/n$ is increasing, so the usual argument for Newton iteration shows that $\mathsf{D}_p(\kappa) = O(\mathsf{M}_p(\kappa))$. Using again that $\mathsf{M}_p(n)/n$ is increasing, we obtain $n\,\mathsf{D}_p(\kappa) = O(\mathsf{M}_p(n\kappa))$.

### 9.2. Multiplication of polynomials over $\mathbb{Z}/p^\alpha\mathbb{Z}$

For any prime $p$ and any integer $\alpha \geqslant 1$, denote by $\mathsf{M}_{p,\alpha}(n)$ the bit complexity of multiplying polynomials in $(\mathbb{Z}/p^\alpha\mathbb{Z})[X]_n$.

Theorem 9.2. *We have*
$$\mathsf{M}_{p,\alpha}(n) = O\bigl(n\alpha \lg p \lg(n\alpha \lg p)\, 8^{\log^*(n\alpha \lg p)}\bigr),$$
*uniformly for all $n, \alpha \geqslant 1$ and all primes $p$. Assuming Conjectures 8.1, 8.3 and 8.5, we may replace $K=8$ by $K=4$.*

For $\lg n = O(\alpha \lg p)$ we may simply use Kronecker substitution. For $\alpha \lg p = O(\lg n)$ we must modify the algorithm of section 7. Recall that in that algorithm we reduced multiplication in $\mathbb{F}_p[X]/(X^n-1)$ to multiplication $\mathbb{F}_{p^\kappa}[Y]/(Y^N-1)$, where $N$ is a transform length dividing $p^\kappa - 1$. Our task is to define a ring $R$, analogous to $\mathbb{F}_{p^\kappa}$, so that multiplication in $(\mathbb{Z}/p^\alpha\mathbb{Z})[X]/(X^n-1)$ can be reduced to multiplication in $R[Y]/(Y^N-1)$.

Let $\pi$ be the natural projection $\mathbb{Z}/p^\alpha \mathbb{Z} \to \mathbb{F}_p$, and write $\pi$ also for the corresponding map $(\mathbb{Z}/p^\alpha \mathbb{Z})[Z] \to \mathbb{F}_p[Z]$. Let $P \in \mathbb{F}_p[Z]$ be any monic irreducible polynomial of degree $\kappa$, and let $\tilde{P} \in (\mathbb{Z}/p^\alpha \mathbb{Z})[Z]$ be an arbitrary lift via $\pi$, also monic of degree $\kappa$. Then $\tilde{P}$ is irreducible, and we will take $R := (\mathbb{Z}/p^\alpha \mathbb{Z})[Z]/\tilde{P}$.

Any primitive $N$-th root of unity $\omega \in \mathbb{F}_{p^\kappa}$ has a unique lift to a principal $N$-th root of unity $\tilde{\omega} \in R$. (This can be seen, for example, by observing that $\tilde{\omega}$ must be the image in $R$ of a Teichmüller lift of $\omega$ in the $p$-adic field whose residue field is $\mathbb{F}_{p^\kappa}$.) Given $\omega$, we may compute $\tilde{\omega}$ efficiently using fast Newton lifting [11, Section 12.3].

Moreover, if we choose $P$ and $\theta \in \mathbb{F}_{p^\kappa} = \mathbb{F}_p[Z]/P$ as in Proposition 6.1, so that $\theta^N = Z$, then fast Newton lifting can also be used to obtain $\tilde{\theta} \in R$ such that $\tilde{\theta}^N = Z$ in $R$.

## 9.3. Multiplication of polynomials over $\mathbb{Z}/m\mathbb{Z}$

For any $m \geqslant 1$, denote by $\mathsf{M}_m(n)$ the bit complexity of multiplying polynomials in $(\mathbb{Z}/m\mathbb{Z})[X]_n$.

THEOREM 9.3. *We have*

$$\mathsf{M}_m(n) = O\bigl(n \lg m \lg (n \lg m) \, 8^{\log^*(n \lg m)}\bigr),$$

*uniformly for all $n, m \geqslant 1$. Assuming Conjectures 8.1, 8.3 and 8.5, we may replace $K = 8$ by $K = 4$.*

We use the isomorphism $\mathbb{Z}/m\mathbb{Z} \cong \prod_i (\mathbb{Z}/p_i^{\alpha_i} \mathbb{Z})$, where $m = p_1^{\alpha_1} \cdots p_l^{\alpha_l}$ is the prime decomposition of $m$. The cost of converting between the $\mathbb{Z}/m\mathbb{Z}$ and $\prod_i (\mathbb{Z}/p_i^{\alpha_i} \mathbb{Z})$ representations is $O(\mathsf{I}(\lg m) \lg l) = O(\mathsf{I}(\lg m) \lg \lg m)$ [17, Section 10.3]. By Theorem 9.2 we get

$$\mathsf{M}_m(n) = O\bigl(n \lg m \lg (n \lg m) \, 8^{\log^*(n \lg m)} + n \, \mathsf{I}(\lg m) \lg \lg m \bigr).$$

The first term dominates if $n \geqslant m$. If $n < m$ we may simply use Kronecker substitution.

## 9.4. Complexity bounds for straight-line programs

Until now we have considered only complexity bounds in the Turing model. The new techniques also lead to improved bounds in algebraic complexity models. In what follows, $\mathcal{A}$ is always a commutative ring with identity.

For the simplest case, first assume that $\mathcal{A}$ is an $\mathbb{F}_p$-algebra for some prime $p$. In the straight-line program model [9, Chapter 4], we count the number of additions and subtractions in $\mathcal{A}$, the number of scalar multiplications (multiplications by elements of $\mathbb{F}_p$), and the number of nonscalar multiplications (multiplications in $\mathcal{A}$). As pointed out in the introduction, the best known bound for the total complexity was previously $O(n \lg n \lg \lg n)$, which was achieved by the Cantor–Kaltofen algorithm.

THEOREM 9.4. *Let $\mathcal{A}$ be an $\mathbb{F}_p$-algebra. We may multiply two polynomials in $\mathcal{A}[X]$ of degree less than $n$ using $O(n \lg n \, 8^{\log^* n})$ additions, subtractions, and scalar multiplications, and $O(n \, 4^{\log^* n})$ nonscalar multiplications. These bounds are uniform over all primes $p$ and all $\mathbb{F}_p$-algebras $\mathcal{A}$.*

The idea of the proof is to use the same algorithm as in section 7, but instead of switching to Kronecker substitution when we reach $n \approx p$, we simply recurse all the way down to $n = 1$. The role of the extension $\mathbb{F}_{p^\kappa}$ is played by $\mathcal{A} \otimes_{\mathbb{F}_p} \mathbb{F}_{p^\kappa} \cong \mathcal{A}[Z]/P$, where $P \in \mathbb{F}_p[Z]$ is monic and irreducible of degree $\kappa$ (here we have used the fact that $\mathbb{F}_p[X]$ may be viewed as a subring of $\mathcal{A}[X]$, since $\mathcal{A}$ contains an identity element and hence a copy of $\mathbb{F}_p$). Thus we reduce multiplication in $\mathcal{A}[X]/(X^n - 1)$ to multiplication in $(\mathcal{A} \otimes_{\mathbb{F}_p} \mathbb{F}_{p^\kappa})[Y]/(Y^N - 1)$. The latter multiplication may be handled by DFTs over $\mathcal{A} \otimes_{\mathbb{F}_p} \mathbb{F}_{p^\kappa}$, since any primitive $N$-th root of unity in $\mathbb{F}_{p^\kappa}$ corresponds naturally to a principal $N$-th root of unity in $\mathcal{A} \otimes_{\mathbb{F}_p} \mathbb{F}_{p^\kappa}$.

The $O(n \lg n \, 8^{\log^* n})$ bound covers the cost of the DFTs, which are accomplished entirely using additions, subtractions and scalar multiplications. Nonscalar multiplications are needed only for the pointwise multiplication step. To explain the $O(n \, 4^{\log^* n})$ bound, we observe that at each recursion level the total "data size" grows by a factor of four: one factor of two arises from the Crandall–Fagin splitting, and another factor of two from the Kronecker substitution.

Note that in the straight-line program model, we give a different algorithm for each $n$. Thus all precomputed objects, such as the defining polynomial $P$ and the required roots of unity, are obtained free of cost (they are built directly into the structure of the algorithm). The uniformity in $p$ follows from the fact that the bounds for $\lambda$ and $M$ in Theorem 4.1 are independent of $p$, although of course the algorithm will be different for each $p$. For each fixed $p$, we may use the same straight-line program for all $\mathbb{F}_p$-algebras $\mathcal{A}$.

Assuming Conjectures 8.3 and 8.5, the bounds may be improved to respectively $O(n \lg n \, 4^{\log^* n})$ and $O(n \, 2^{\log^* n})$. However, in this case we lose uniformity in $p$, due to the issue raised in Remark 8.6.

Theorem 9.4 can be generalised to $(\mathbb{Z}/p^\alpha \mathbb{Z})$-algebras, along the same lines as section 9.2. It is also possible to handle $(\mathbb{Z}/m\mathbb{Z})$-algebras for any integer $m \geqslant 1$, i.e., rings of finite characteristic, but we cannot proceed exactly as in section 9.3, because the straight-line model has no provision for a "reduction modulo $p^\alpha$" operation. Instead, we use a device introduced in [10]. Suppose that $m = p_1^{\alpha_1} \cdots p_l^{\alpha_l}$ and that $\mathcal{A}$ is now a $(\mathbb{Z}/m\mathbb{Z})$-algebra. For each $i$ we may construct a straight-line program $\mathcal{M}_i$ that takes as input polynomials $f, g \in \mathcal{B}[X]$ of degree less than $n$, where $\mathcal{B}$ is any $(\mathbb{Z}/p_i^{\alpha_i}\mathbb{Z})$-algebra, and computes $fg$. By replacing every constant in $\mathbb{Z}/p_i^{\alpha_i}\mathbb{Z}$ by a compatible constant in $\mathbb{Z}/m\mathbb{Z}$, we obtain a straight-line program $\mathcal{M}_i'$ that takes as input $f, g \in \mathcal{A}[X]$ of degree less than $n$, and computes $h_i \in \mathcal{A}[X]$ such that $h_i - fg \in p_i^{\alpha_i}\mathcal{A}[X]$. By the Chinese remainder theorem we may choose $e_i \in \mathbb{Z}/m\mathbb{Z}$ such that $e_i = 0 \pmod{m/p_i^{\alpha_i}}$ and $e_i = 1 \pmod{p_i^{\alpha_i}}$ for each $i$. The linear combination $\sum_i e_i h_i$ is then equal to $fg$ in $\mathcal{A}[X]$. We conclude that we may multiply polynomials in $\mathcal{A}[X]$ using $O(n \lg n \, 8^{\log^* n})$ additions, subtractions and scalar multiplications (by elements of $\mathbb{Z}/m\mathbb{Z}$), and $O(n \, 4^{\log^* n})$ nonscalar multiplications (i.e., multiplications in $\mathcal{A}$). These bounds are not uniform in $m$, but for each $m$ they are uniform over all $(\mathbb{Z}/m\mathbb{Z})$-algebras.

## 9.5. Other algebraic models

If we wish to take into account the cost of precomputations, and give a single algorithm that works uniformly for all $p$ and $n$, we may use a more refined complexity model such as the BSS model [6]. In this model, a *machine over $\mathcal{A}$* is, roughly speaking, a Turing machine in which the tape cells hold elements of $\mathcal{A}$. Actually, we need a multi-tape version of the model described in [6]. The machine can perform arithmetic operations on elements of $\mathcal{A}$ in unit time, but can also manipulate data such as index variables in the same way as a Turing machine, and must deal with data layout in the same way as a Turing machine. In this model we obtain similar bounds to Theorem 9.4, but without uniformity in $p$. Alternatively, we could obtain bounds uniform in $p$ if we added extra terms to account for the precomputations.

Another point of view in Theorem 9.4 is that we have described a new *evaluation-interpolation strategy* for polynomials over an $\mathbb{F}_p$-algebra $\mathcal{A}$. We refer the reader to [21, Sections 2.1–2.4] for classical examples of evaluation-interpolation schemes, and [16] for algorithms specific to finite fields. Such schemes are characterised by two quantities: the evaluation/interpolation complexity $\mathsf{E}(n)$ and the number $\mathsf{N}(n)$ of evaluation points. The new algorithms yield the bounds $\mathsf{E}(n) = O(n \lg n \, 8^{\log^* n})$ and $\mathsf{N}(n) = O(n \, 4^{\log^* n})$, and these can be used to prove complexity bounds for problems more general than polynomial multiplication. For example, we can multiply $r \times r$ matrices with entries in $\mathcal{A}[X]_n$ using $O(r^2 n \lg n \, 8^{\log^* n} + r^\omega n \, 4^{\log^* n})$ ring operations, where $\omega$ is an exponent of matrix multiplication. One of the main advantages of our algorithms is that $\mathsf{N}(n)$ is almost linear, contrary to synthetic FFT methods [10, 31] derived from Schönhage–Strassen multiplication [32], which achieve only $\mathsf{N}(n) = O(n \lg n)$.

In the setting of bilinear complexity [9, Chapter 14], the new algorithms do not improve asymptotically on the best known bounds. For example, it is known [27] that for any $n$ there exist $\mathbb{F}_2$-linear maps $a_i, b_i \colon \mathbb{F}_2[X]_n \to \mathbb{F}_2$ and polynomials $c_i \in \mathbb{F}_2[X]_{2n}$ for $i \in \{1, \ldots, k(n)\}$, with $k(n) = (189/22 + o(1)) \, n$, such that $u\,v = \sum_{i=1}^{k(n)} a_i(u) \, b_i(v) \, c_i$ for all $u, v \in \mathbb{F}_2[X]_n$. The new method yields the asymptotically inferior bound $k(n) = O(n \, 4^{\log^* n})$. Nevertheless, these bounds are asymptotic and do not take into account that our new algorithm would rather work over (say) $\mathbb{F}_{2^{60}}$ instead of $\mathbb{F}_2$. In practice, the new method might therefore outperform the bilinear algorithms from [27].

## 9.6. Implementations

The new complexity bounds have a quite theoretic flavour: in practice, slowly growing functions such as $\log \log n$ and $8^{\log^* n}$ really behave as constants. Furthermore, for concrete applications, it would be more relevant to redefine $\log^*$ as

$$\log^* n \ := \ \min \{k \in \mathbb{N} : \log^{\circ k} n \leqslant 64\}.$$

With this definition, we observe that $\log^* n \leqslant 1$ for all feasible values of $n$ in the foreseeable future. Nevertheless, we think that variants of the ideas in this paper could still be useful in practice.

Let us briefly sketch how an efficient multiplication algorithm over $\mathbb{F}_2$ could be implemented. Various modern processors offer a special instruction for the multiplication of two polynomials in $\mathbb{F}_2[X]_{64}$. The practical translation of the results from section 4 is to work over the field $\mathbb{F}_{2^{60}} = \mathbb{F}_2[Z]/P$ where $P := Z^{60} + \cdots + Z + 1$. This has two advantages. First, if we represent elements of $\mathbb{F}_{2^{60}}$ by their residues modulo $Z^{61} - 1$, i.e., using a redundant representation, then each multiplication in $\mathbb{F}_{2^{60}}$ requires a single hardware multiply instruction, followed by simple shift and XOR instructions. Second, there exist primitive roots of unity of high smooth order $N = 2^{60} - 1 = 3^2 \cdot 5^2 \cdot 7 \cdot 11 \cdot 13 \cdot 31 \cdot 41 \cdot 61 \cdot 151 \cdot 331 \cdot 1321$.

Transforms of small prime lengths $p \mid N$ over $\mathbb{F}_{2^{60}}$ can be implemented using specialised codelets, similar to those implemented in FFTW3 [14]. Roughly speaking, the constant factor of a radix-$p$ DFT is $p/\log p$. In our case, $(13/\log 13)/(2/\log 2) \approx 1.8$, so specialised codelets should remain reasonably efficient for $p \leqslant 13$. Consequently, through the mere use of such small radices, we obtain efficient algorithms for computing DFTs of lengths dividing $3^2 \cdot 5^2 \cdot 7 \cdot 11 \cdot 13 = 225225$. To handle larger primes, we may use any algorithm for converting DFTs to convolutions. Rather than use Bluestein's algorithm as in the theoretical portion of the paper, it is more economical in this case to use Rader's algorithm, which reduces a DFT of length $p$ to a cyclic convolution of length $p-1$, together with $O(p)$ additions in $\mathbb{F}_{2^{60}}$ (see also Remark 2.4).

More precisely, consider the remaining transform lengths $p = 31, 41, 61, 151, 331$ and $1321$. We have $30 = 2 \cdot 3 \cdot 5$, $40 = 2^3 \cdot 5$, $60 = 2^2 \cdot 3 \cdot 5$, $150 = 2 \cdot 3 \cdot 5^2$, $330 = 2 \cdot 3 \cdot 5 \cdot 11$ and $1320 = 2^3 \cdot 3 \cdot 5 \cdot 11$, so in each case, the DFT reduces to a two-dimensional convolution of size $2^i \times d$, where $d$ is a divisor of $225225$ and $1 \leqslant i \leqslant 3$. Such a convolution in turn reduces to $2^{i+1}$ DFTs of length $d$ together with $d$ convolutions of length $2^i$. The latter cannot be handled using DFTs over $\mathbb{F}_{2^{60}}$, since there are no roots of unity of order 2, 4 or 8, but in any case they are all very short convolutions and could be performed directly, or by using Karatsuba's algorithm.

Transforms of other lengths dividing $N$ may be reduced to these cases via the Cooley–Tukey algorithm. Finally, to handle arbitrary input lengths efficiently, we may use a mixed-radix generalisation of the truncated Fourier transform [19, 20].

We also reemphasise the fact that DFTs of this kind can be used as an evaluation-interpolation technique. This approach is for instance attractive for the multiplication of polynomial matrices over $\mathbb{F}_2$.

We have not yet implemented any of these algorithms. We expect that much fine-tuning will be necessary to make them most effective in practice. We intend to report on this issue in a future work.